\documentclass[a4paper,UKenglish]{lipics-v2016}

\usepackage{tabularx,colortbl}
\usepackage[table]{xcolor}
\usepackage{graphicx}
\usepackage{stmaryrd}
\usepackage{mathtools}
\usepackage{appendix}

\makeatletter
\newif\if@restonecol
\makeatother

\usepackage[ruled,noend,linesnumbered,lined]{algorithm2e}

\definecolor{gold}{rgb}{0.99,0.78,0.07}

\usepackage{algpseudocode}
\usepackage{xypic}
\usepackage{enumerate}

\usepackage{ifthen} 
\usepackage{xifthen} 

\usepackage{verbatim}

\newcommand{\THIN}{\mathrm{Thin}}
\newcommand{\THICK}{\mathrm{Thick}}
\def\rmdef{\stackrel{\mbox{\rm {\tiny def}}}{=}} 
\newcommand{\ttime}{\mathsf{time}}
\newcommand{\sinf}{\mbox{\sf {\tiny inf}}}
\newcommand{\ssup}{\mbox{\sf {\tiny sup}}}

\newcommand{\Opt}{\text{Opt}}
\newcommand{\mMIN}{\text{Min}}
\newcommand{\mMAX}{\text{Max}}
\newcommand{\mMin}{\text{Min}}
\newcommand{\mMax}{\text{Max}}

\newcommand{\set}[1]{\{ #1 \}}
\newcommand{\seq}[1]{\langle #1 \rangle}

\newcommand{\Aa}{{\cal A}}

\newcommand{\Rr}{{\cal R}}

\newcommand{\Gg}{{\cal G}}

\newcommand{\NATS}[1]{\llbracket #1 \rrbracket_\Nat}

\newcommand{\CC}{\mathrm{CC}}
\newcommand{\zones}{{\cal Z}}
\newcommand{\CLOS}[1]{\overline{#1}}

\newcommand{\act}{\mathit{Act}}

\newcommand{\Gp}{G^\boxplus}
\newcommand{\Bp}{B^\boxplus}

\newcommand{\RegB}[1]{\widehat{#1}}
\newcommand{\hSigma}{\RegB{\Sigma}}
\newcommand{\hPi}{\RegB{\Pi}}

\newcommand{\hpta}{\RegB{\pta}}
\newcommand{\hS}{\RegB{S}}
\newcommand{\hA}{\RegB{A}}

\newcommand{\hpi}{\RegB{\pi}}
\newcommand{\hT}{\RegB{T}}

\newcommand{\hs}{\hat{s}}

\newcommand{\sat}{\in}

\newcommand{\FRUNS}{\text{Runs}_{\text{fin}}}
\newcommand{\RUNS}{\text{Runs}}

\newcommand{\LAST}{\text{Last}}

\newcommand{\RUN}{\text{Run}}
\newcommand{\VAL}{\text{Val}}

\usepackage{amsmath,amsfonts,amssymb,xspace}
\usepackage{wasysym} 
\usepackage{color}

\usepackage[inline]{enumitem}
\setlist{nosep}

\usepackage{array}
\newcolumntype{x}[1]{>{\centering\let\newline\\\arraybackslash\hspace{0pt}}p{#1}}

\usepackage{tikz}

\usetikzlibrary{arrows,shapes,snakes,automata,backgrounds,positioning,decorations.pathmorphing,
  decorations.markings,calc}

\tikzstyle{player1}=[draw=black,rounded rectangle, minimum size=5mm,fill=gold]
\tikzstyle{player2}=[draw=black,rectangle,minimum size=5mm,fill=gold]

\tikzstyle{widget}=[draw=gold,rectangle, rounded rectangle=10pt,dashed,minimum size=6mm,fill=gold!20]
\tikzset{every loop/.style={looseness=7}}
\tikzset{
    gluon/.style={decorate,draw=black,
        decoration={coil,amplitude=1pt, segment length=5pt}} 
}
\tikzset{
    gluon1/.style={decorate,draw=black,
        decoration={coil,amplitude=3pt, segment length=3pt}} 
}
\tikzset{
    gluonew/.style={decorate,draw=black,
        decoration={coil,amplitude=1pt, segment length=2pt}} 
}

\renewcommand\geq{\geqslant}
\renewcommand\leq{\leqslant}

\newcommand\clocks{\mathcal X}

\newcommand\Nat{\mathbb{N}}

\newcommand\Real{\mathbb{R}}
\newcommand\Rplus{\mathbb{R}_{\geq 0}}

\newcommand\N{\Nat}

\newcommand{\argmaxlex}{\operatornamewithlimits{argmax^{\mathrm{lex}}}}
 
\newcommand{\argmax}{\operatornamewithlimits{argmax}}
\newcommand{\argmin}{\operatornamewithlimits{argmin}}

\newcommand\location{\ell}

\newcommand{\sem}[1]{ [ \! [ {#1}  ]  \! ]} 

\newcommand{\trans}[1]{\xrightarrow{#1}}

\newcommand\REACH{\textsf{Reach}}
\newcommand\RCPS{\textsf{RCPS}}
\newcommand\RCPSs{\textsf{RCPSs}}

\newcommand{\region}{\zeta}

\renewcommand\paragraph[1]{\noindent \textbf{#1.}}

\newcommand{\pta}{{\mathsf{T}}}
\newcommand{\MPG}{{\mathsf{MPG}}}
\newcommand{\inv}{\mathit{Inv}}

\colorlet{AlgCaptionColor}{gold!80}

\makeatletter
\renewcommand{\algocf@makecaption@ruled}[2]{%
  \global\sbox\algocf@capbox{\colorbox{AlgCaptionColor}{\hskip\AlCapHSkip
      \parbox[t]{\hsize}{\algocf@captiontext{\strut#1}{\strut#2\strut}}\hskip.6\algomargin}}
      }%
      \makeatother      

\title{Mean-Payoff Games on Timed Automata}

\author[1]{S. Guha}
\author[2]{M. Jurdzi{\'n}ski}
\author[3]{S. N. Krishna}
\author[4]{A. Trivedi}
\affil[1]{
   Indian Institute of Technology  Delhi, India (\texttt{shibashis@cse.iitd.ac.in})} 
\affil[2]{
  The University of Warwick, UK (\texttt{marcin@dcs.warwick.ac.uk})}
\affil[3]{
  Indian Institute of Technology  Bombay, India (\texttt{krishnas@cse.iitb.ac.in})} 
\affil[4]{
  University of Colorado Boulder, USA (\texttt{ashutosh.trivedi@colorado.edu})} 

\authorrunning{Guha, Jurdzi{\'n}ski, Krishna, and Trivedi}
\Copyright{Guha, Jurdzi{\'n}ski, Krishna, and Trivedi}


\begin{document}
\maketitle

\begin{abstract}
Mean-payoff games on timed automata are played on the infinite weighted graph of 
configurations of priced timed automata between two players---Player Min and Player
Max---by moving a token along the states of the graph to form an infinite run.  
The goal of Player Min is to minimize the limit average weight of the run, while
the goal of the Player Max is the opposite.  
Brenguier, Cassez, and Raskin recently studied a variation of these games and
showed that mean-payoff games are undecidable for timed automata with 
five or more clocks.  
We refine this result by proving the undecidability of
mean-payoff games with three clocks. 
On a positive side, we show the decidability of mean-payoff games on one-clock
timed automata with binary price-rates.  
A key contribution of this paper is the application of dynamic programming
based proof techniques applied in the context of average reward optimization
on an uncountable state and action space. 
\end{abstract}

\section{Introduction}
\label{sec:intro}
The classical mean-payoff
games~\cite{ZP96,EM79,GKK88,BSV04} are 
two-player zero-sum games that are played on weighted finite graphs, where two
players---Max and Min---take turn to move a token along the edges of the graph
to jointly construct an infinite play.
The objectives of the players Max and Min are to respectively maximize
and minimize the limit average reward associated with the play.  
Mean-payoff games are well-studied in the context of optimal controller
synthesis in the framework of Ramadge-Wonham~\cite{RW89}, where the goal of the
game is to find a control strategy that maximises the average reward earned
during the evolution of the system.    
Mean-payoff games enjoy a special status in verification, since $\mu$-calculus
model checking and parity games can be reduced in polynomial-time to solving
mean-payoff games.   
Mean-payoff objectives can also be considered as quantitative
extensions~\cite{Henzinger2013} of 
classical B\"uchi objectives, where we are interested in the limit-average share of
occurrences of accepting states rather than merely in whether or not infinitely
many accepting states occur.  
For a broader discussion on quantitative verification, in general, and the transition
from the classical qualitative to the modern quantitative interpretation of
deterministic B\"uchi automata, we refer the reader to Henzinger's excellent survey~\cite{Henzinger2013}. 

We study mean-payoff games played on an infinite configuration graph of timed
automata. 
Asarin and Maler~\cite{AM99} were the first  to study games on timed automata
and they gave an algorithm to solve timed games with reachability time objective.
Their work was later generalized and improved upon by Alur et
al.~\cite{ABM04} and Bouyer et al.~\cite{BCFL04}.
Bouyer et al.~\cite{BBL04,Bou06} also studied the more difficult
average payoffs, but only in the context of scheduling, which in
game-theoretic terminology corresponds to 1-player games.
However, they left the problem of proving decidability of 2-player average
reward games on priced timed automata open.
Jurdzi\'nski and Trivedi~\cite{JT08b} proved the decidability of the special
case of average time games  where all locations have unit costs.
More recently, mean-payoff games on timed automata have been studied by
Brenguier, Cassez and Raskin \cite{BCR14} where they consider average payoff per
time-unit.
Using the undecidability of \emph{energy games} \cite{BLM12}, they showed
undecidability of mean-payoff games on weighted timed games with five or more clocks.
They also gave a semi-algorithm to solve cycle-forming games on timed automata
and characterized the conditions under which a solution of these games gives a
solution for mean-payoff games.

On the positive side, we characterize general conditions under which dynamic
programming based techniques can be used to solve the mean-payoff games on timed
automata.
As a proof-of-concept, we consider one-clock binary-priced timed games, and prove
the decidability of mean-payoff games for this subclass.
Our decidability result can be considered as the average-payoff analog of the
decidability result by Brihaye et al.~\cite{BGN+14} for reachability-price games
on timed automata.
We strengthen the known undecidability results for mean-payoff
games on timed automata in three ways:
i) we show that the mean-payoff games over priced timed games is undecidable
for timed games with only three clocks;
ii) secondly, we show that undecidability can be achieved with binary
price-rates;
and finally, iii) our undecidability results are applicable for problems where
the average payoff is considered per move as well as for problems when it is
defined per time-unit.

Howard~\cite{Howard60,Put94} introduced gain and bias optimality equations to
characterize optimal average on one-player finite game arenas.
Gain and bias optimality equations based characterization has been extended to
two-player game arenas~\cite{FV97} as well as many subclasses of uncountable
state and action spaces~\cite{DY79,BBJLR08}.
The work of Bouyer et al.~\cite{BBJLR08} is perhaps the closest to our
approach---they extended optimality equations approach to solve games on hybrid
automata with certain strong reset assumption that requires all continuous
variables to be reset at each transition, which in the case of timed automata is
akin to requiring all clocks to be reset at each transition.  
To the best of our knowledge, the exact decidability for timed games does not
immediately follow from any previously known results. 

Howard's Optimality equations requires two variable per state: the gain of the
state and the bias of the state. 
Informally speaking, the gain of a state corresponds to the optimal mean-payoff for
games starting from that state, while the bias corresponds to the limit of
transient sum of step-wise deviations from the optimal average.
Hence, intuitively at a given point in a game, both players would prefer to first
optimize the gain, and then choose to optimize bias among choices with equal gains. 
We give general conditions under which a solution of gain-bias
equations for a finitary abstraction of timed games can provide a solution of
gain-bias equations for the original timed game.
For this purpose, we exploit a region-graph like abstraction of timed
automata~\cite{JT07} called the
boundary region abstraction (BRA).
Our key contribution is the theorem that states that every solution of gain-bias
optimality equations for boundary region abstraction carries over to the
original timed game, as long as for every region, the gain
values are constant and the bias values are affine. 

The paper is organized in the following manner. 
In Section~\ref{sec:prelims} we describe  mean-payoff games
and introduce the notions of \emph{gain} and \emph{bias} optimality equations.
This section also introduces mean-payoff games over timed automata and states
the key results of the paper.
Section \ref{sec:brg} introduces the \emph{boundary region abstraction} 
for timed automata and characterizes the conditions under which the solution of a
game played over the boundary region abstraction can be lifted to a solution of
mean payoff game over priced timed automata. 
In Section~\ref{sec:dec} we present the  strategy improvement algorithm
to solve optimality equations for mean-payoff games played over boundary region
abstraction and connect them to solution of optimality equations over
corresponding timed automata. 
Finally, Section \ref{sec:undec} sketches the undecidability of mean-payoff
games for binary-priced timed automata with three clocks.

\section{Mean-Payoff Games on Timed Automata}
\label{sec:prelims}
We begin this section by introducing mean-payoff games on graphs with
uncountably infinite vertices and edges, and show how, and under
what conditions, gain-bias optimality equations characterize the value of mean-payoff
games.
We then set-up mean-payoff games for timed automata and state our key
contributions. 
\subsection{Mean-Payoff Games}
\label{defn:mpg}
\begin{definition}[Turn-Based Game Arena]
  A game arena $\Gamma$ is a tuple 
  $(S, S_\mMIN, S_\mMAX,  A, T, \pi)$ where 
    $S$ is a (potentially uncountable) set of states partitioned between sets
    $S_\mMIN$ and $S_\mMAX$ of states controlled by Player Min and Player Max,
    respectively; 
    $A$ is a (potentially uncountable) set of \emph{actions};
    $T: S \times A \to S$ is a partial function called  \emph{transition
    function};  and 
    $\pi: S \times A \to \Real$ is a partial function called \emph{price function}.  
\end{definition}

We say that a game arena is \emph{finite} if both $S$ and $A$ are finite. 
For any state $s \in S$, we let $A(s)$ denote the set of actions available in
$s$, i.e.,  the actions $a \in A$ for which $T(s, a)$ and $\pi(s, a)$ are
defined. 
A transition of a game arena is a tuple $(s, a, s') \in S {\times} A {\times} S$
such that $s' = T(s, a)$ and we write $s \xrightarrow{a} s'$. 
A finite play starting at a state $s_0$ is a
sequence of transitions  $\seq{s_0, a_1, s_1, a_2, \ldots, s_n} \in S {\times}
(A {\times} S)^*$ 
such that for all $0 \leq i < n$ we have that 
$s_i \xrightarrow{a_{i+1}} s_{i+1}$ is a transition.  
For a finite play $\rho = \seq{s_0, a_1, \ldots, s_n}$ we write
$\LAST(\rho)$ for the final state of $\rho$, here $\LAST(\rho) = s_n$. 
The concept of an infinite play $\seq{s_0, a_1, s_1, \ldots}$ is defined in an
analogous way.  
We write $\RUNS(s)$ and $\FRUNS(s)$ for the set of plays and
the set of finite plays starting at $s \in S$ respectively. 

A \emph{strategy} of Player $\mMIN$ is a function $\mu : \FRUNS \to A$
such that $\mu(\rho) \in A(\LAST(\rho))$ for all finite plays $\rho \in \FRUNS$,
i.e.\ for any finite play, a strategy of $\mMIN$ returns an action
available to $\mMIN$ in the last state of the play. 
A strategy $\chi$ of $\mMAX$ is defined analogously and we let $\Sigma_\mMIN$
and $\Sigma_\mMAX$ denote the sets of strategies of $\mMIN$ and $\mMAX$,
respectively. 
A strategy $\sigma$ is \emph{positional} if 
$\LAST(\rho) {=} \LAST(\rho')$ implies $\sigma(\rho) {=} \sigma(\rho')$ for all  
$\rho, \rho' \in \FRUNS$. 
This allows us to represent a positional strategy as a function in $[S \to A]$.
Let  $\Pi_\mMIN$ and $\Pi_\mMAX$ denote the set of positional strategies of
$\mMIN$ and $\mMAX$, respectively. 
For any state $s$ and strategy pair $(\mu,\chi) \in  \Sigma_\mMIN {\times}
\Sigma_\mMAX$, let $\RUN(s, \mu, \chi)$
denote the unique infinite play  
$\seq{s_0, a_1, s_1, \ldots}$ in which  $\mMIN$ and $\mMAX$ play according to
$\mu$ and $\chi$, respectively, i.e. 
for all $i \geq 0$ we have that 
$s_i \in S_\mMIN$ implies $a_{i+1} = \mu(\seq{s_0, a_1, \ldots, s_i})$ and 
$s_i \in S_\mMAX$ implies $a_{i+1} = \chi(\seq{s_0, a_1, \ldots, s_i})$.

In a mean-payoff game on a game arena, players $\mMIN$ and $\mMAX$ move a token
along the transitions indefinitely thus forming an infinite play 
$\rho = \seq{s_0, a_1, s_1, \ldots}$ in the game graph.  
The goal of player $\mMIN$ is to minimize 
$\Aa_\mMIN(\rho) = 
\limsup_{n \to \infty} \frac{1}{n} \cdot \sum_{i=0}^{n-1} \pi(s_i, a_{i+1})$
and  
the goal of player $\mMAX$ is to maximize 
$\Aa_\mMAX(\rho) = 
\liminf_{n \to \infty} \frac{1}{n}\cdot \sum_{i=0}^{n-1} \pi(s_i, a_{i+1})$.
The \emph{upper value} $\VAL^*(s)$ and the lower value $\VAL_*(s)$ of a state
$s \in S$ are defined as: 
\[
\VAL^*(s) = \inf_{\mu \in \Sigma_\mMIN} \sup_{\chi \in \Sigma_\mMAX}
\Aa_\mMIN(\RUN(s, \mu, \chi)) \text{ and }
\VAL_*(s) = \sup_{\chi \in \Sigma_\mMAX} \inf_{\mu \in \Sigma_\mMIN}
\Aa_\mMAX(\RUN(s, \mu, \chi))
\]
respectively.
It is always the case that $\VAL_*(s) \leq \VAL^*(s)$. 
A mean-payoff game is called \emph{determined} if for every state $s \in S$
we have that $\VAL_*(s) = \VAL^*(s)$. 
Then, we write $\VAL(s)$ for this number and we call it the \emph{value} of the
mean-payoff game at state~$s$.
We say that a game is \emph{positionally-determined} if for every
$\varepsilon > 0$ we have strategies $\mu_\varepsilon \in \Pi_\mMIN$ and
$\chi_\varepsilon \in \Pi_\mMAX$
such that for every initial state $s \in S$, we have that 
\[
\VAL_*(s) - \varepsilon  \leq
  \inf_{\mu' \in \Sigma_\mMIN} \Aa_\mMAX(\RUN(s, \mu', \chi_\varepsilon))\text{ and }
  \VAL^*(s) + \varepsilon  \geq   
  \sup_{\chi' \in \Sigma_\mMAX} \Aa_\mMIN(\RUN(s, \mu_\varepsilon, \chi')).
\]
For a given $\varepsilon$ we call each such strategy an $\varepsilon$-optimal
strategy for the respective player.


Given two functions $G: S \to \Real$ (gain) and $B: S \to \Real$ (bias), we say that $(G, B)$
is a solution to the optimality equations for  mean-payoff game on 
$\Gamma = (S, S_\mMIN, S_\mMAX, A, T, \pi)$, denoted $(G, B) \models
\Opt(\Gamma)$  if 
\begin{eqnarray*}
  G(s) & = & 
  \begin{cases}
    \sup_{a \in A(s)}  \set{G(s') \::\: s \xrightarrow{a} s' } &  \text{ if $s \in S_\mMAX$}\\
    \inf_{a \in A(s)} \set{G(s') \::\: s \xrightarrow{a} s' } &   \text{ if $s \in S_\mMIN$}.
  \end{cases}\\
  B(s) & = & 
  \begin{cases}
    \sup_{a \in A(s)}  \set{ \pi(s, a) - G(s) + B(s') \::\: s \xrightarrow{a} s'
      \text{ and } G(s) = G(s')} &  \text{ if $s \in S_\mMAX$}\\
    \inf_{a \in A(s)} \set{\pi(s, a) - G(s) + B(s') \: : \: s \xrightarrow{a} s'
      \text{ and } G(s) = G(s')} &   \text{ if $s \in S_\mMIN$}.
  \end{cases} 
\end{eqnarray*}
We prove the following theorem connecting a solution of the optimality equations
with mean-payoff games.
We exploit this theorem to solve mean-payoff games on timed automata.
\begin{theorem}
\label{thm:gain-bias-correctness}
  If there exists a function $G: S \to \Real$ with finite image and a function
  $B: S \to \Real$ with bounded image such that $(G, B) \models \Opt(\Gamma)$ then
  for every state  $s \in S$, we have that $G(s) = \VAL(s)$ and for every
  $\varepsilon >0$ both players have positional $\varepsilon$-optimal strategies.
\end{theorem}
\begin{proof}
 Assume that we are given the functions $G: S \to \Real$ with finite image and
 $B: S \to \Real$ with bounded image such that $(G, B) \models \Opt(\Gamma)$. 
 In order to prove the result we show, for every $\varepsilon > 0$, the
 existence of positional strategies $\mu_\varepsilon$ and $\chi_\varepsilon$ such that 
 \begin{eqnarray*}
   G(s) - \varepsilon  \leq 
   \inf_{\mu' \in \Sigma_\mMIN} \Aa_\mMAX(\RUN(s, \mu', \chi_\varepsilon))\text{ and } 
   G(s) + \varepsilon  \geq 
   \sup_{\chi' \in \Sigma_\mMAX} \Aa_\mMIN(\RUN(s, \mu_\varepsilon, \chi')).
 \end{eqnarray*}
 The proof is in two parts.
 \begin{itemize}
 \item 
   Given $\varepsilon > 0$ we compute the positional strategy $\mu_\varepsilon \in
   \Pi_\mMIN$ satisfying the following conditions: 
   $\mu_\varepsilon(s) = a$ if 
   \begin{eqnarray}
     G(s) &=& G(s') \\
     B(s) &\geq& \pi(s, a) - G(s) + B(s') - \varepsilon,
   \end{eqnarray}
   where $s \xrightarrow{a} s'$.
   Notice that it is always possible to find such strategy since $(G, B)$ satisfies
   optimality equations and $G$ is finite image.
   
   Now consider an arbitrary strategy $\chi \in \Sigma_\mMAX$ and consider the run
   $\RUN(s, \mu_\varepsilon, \chi) = \seq{s_0, a_1, s_1, \ldots, s_n, \ldots}$. 
   Notice that for every $i \geq 0$ we have that $G(s_i) \geq G(s_{i+1})$ if 
   $s_i \in S_\mMAX$ and $G(s_i) = G(s_{i+1})$ if $s_i \in S_\mMIN$. 
   Hence $G(s_0), G(s_1), \ldots$ is a non-increasing sequence. 
   Since $G$ is finite image, the sequence eventually becomes constant. 
   Assume that for $i \geq N$ we have that $G(s_i) = g$.
   Now notice that for all $i \geq N$ we have that $B(s_i) \geq \pi(s_i, a_{i+1}) - g +
   B(s_{i+1})$ if $s_i \in S_\mMAX$ and $B(s_i) \geq \pi(s_i, a_{i+1}) - g + B(s_{i+1})
   - \varepsilon$ if $s_i \in S_\mMIN$. 
   Summing these equations sidewise form $i=N$ to $N+k$ we have that 
   $B(s_N) \geq \sum_{i=N}^{N+k} \pi(s_i, a_{i+1}) - (k+1)\cdot g + B(s_{N+k+1}) -
   (k+1)\cdot \varepsilon$.
   Rearranging, we get 
   \begin{eqnarray*}
     g  &\geq& \frac{1}{k+1} \sum_{i=N}^{N+k} \pi(s_i, a_{i+1}) +  \frac{1}{k+1}
     (B(s_{N+k+1}) - B(s_N)) - \varepsilon. \\
     \text{Hence } g &\geq& \limsup_{k\to\infty} \frac{1}{k+1} \sum_{i=N}^{N+k} \pi(s_i, a_{i+1}) +
     \limsup_{k \to \infty}  \frac{1}{k+1} (B(s_{N+k+1}) - B(s_N)) - \varepsilon \\
     &= & \limsup_{k\to\infty} \frac{1}{k} \sum_{i=0}^{k} \pi(s_i, a_{i+1}) - \varepsilon \\
     \text{Hence }  G(s) + \varepsilon & \geq& \Aa_\mMIN(\RUN(s, \mu_\varepsilon, \chi)).
   \end{eqnarray*}
   Since $\chi$ is an arbitrary strategy in $\Sigma_\mMAX$,
   we have $G(s) + \varepsilon \geq 
   \sup_{\chi' \in \Sigma_\mMAX} \Aa_\mMIN(\RUN(s, \mu_\varepsilon, \chi'))$.
 \item
   This part is analogous to the first part of the proof and is omitted.
 \end{itemize}
 The proof is now complete.
\end{proof}

\subsection{Timed Automata}
Priced Timed Game Arenas (PTGAs) extend classical timed
automata~\cite{AD94} with a partition of the actions between two players $\mMin$ and $\mMax$. 
Before we present the syntax and semantics of PTGAs, we need to introduce
the concept of clock variables and related notions. 

\noindent{\bf Clocks.}
Let $\clocks$ be a finite set of \emph{clocks}.
A \emph{clock valuation} on $\clocks$ is a function
$\nu : \clocks {\to} \Rplus$ and we write $V(\clocks)$ (or just $V$ when
$\clocks$ is clear from the context) for the set of clock valuations.
Abusing notation, we also treat a valuation $\nu$ as a point in
$(\Rplus)^{|\clocks|}$. Let $\mathbf{0}$ denote the clock valuation that assigns
0 to all clocks. 
If~$\nu \in V$ and $t \in \Rplus$ then we write $\nu {+} t$ for the
clock valuation defined by $(\nu {+} t)(c) = \nu(c) {+} t$ for all 
$c \in \clocks$.
For $C \subseteq \clocks$, we write $\nu_C$ for the valuation
where $\nu_C(c)$ equals $0$ if 
$c \in C$ and $\nu(c)$ otherwise.
For $X \subseteq V(\clocks)$, we write $\CLOS{X}$ for the smallest closed set
in~$V$ containing $X$.  
Although clocks are usually allowed to take arbitrary non-negative values,
for notational convenience we assume that there is a $K \in \Nat$ such that
for every $c \in \clocks$ we have $\nu(c) \leq K$.

\noindent{\bf Clock Constraints.}
A \emph{clock constraint} over $\clocks$ with upper bound $K \in \Nat$ is a
conjunction of \emph{simple constraints} of the form $c \bowtie i$ or $c {-} c'
\bowtie i$, where  $c, c' \in \clocks$, $i \in \Nat$, $i {\leq} K$, and 
${\bowtie} \in \{{<}, {>}, {=}, {\leq} , {\geq} \}$. 
For $\nu \in V(\clocks)$ and $K \in \Nat$, let $\CC(\nu, K)$ be the set of clock constraints with
upper bound $K$ which hold in~$\nu$, i.e.\ those constraints that resolve to
$\mathtt{true}$ after substituting each occurrence of a clock $x$ with $\nu(x)$.

\noindent{\bf Regions and Zones.}
Every clock region is an equivalence class of the
indistinguishability-by-clock-constraints relation.   
For a given set of clocks $\clocks$ and upper bound $K \in \Nat$ on
clock constraints, a \emph{clock
  region} is a maximal set $\region {\subseteq} V(\clocks)$ such that 
$\CC(\nu, K) {=} \CC(\nu', K)$ for all $\nu, \nu' \in \region$. 
For the set of clocks $\clocks$ and upper bound $K$ we write
$\Rr(\clocks, K)$ for the corresponding finite set of clock regions. 
We write $[\nu]$ for the clock region of $\nu$. 
A \emph{clock zone} is a convex set of clock valuations that satisfies constraints of the form
$\gamma ::= \; c_1 \bowtie k \:\vert\: c_1-c_2 \bowtie k \:\vert\: \gamma \wedge \gamma$,
$k \in \N$, $c_1,c_2 \in \clocks$ and $\bowtie \: \in \: \{\le, <, =, >, \ge\}$.
We write $\zones(\clocks,K)$ for the set of clock zones over the set of clocks
$\clocks$ and upper bound $K$. 
When $\clocks$ and $K$ are clear from the context we write  $\Rr$
and $\zones$ for the set of regions and zones.
In this paper we fix a positive integer $K$, and work with
$K$-bounded clocks and clock constraints. 

%
%
\subsection{Priced Timed Game Arena: Syntax and Semantics}
\begin{definition}
  A priced timed game arena is a tuple 
  $\pta {=} (L_\mMIN, L_\mMAX, \act, \clocks, \inv,  E, \rho, \delta, p)$
  where 
    $L_\mMIN$ and $L_\mMAX$ are sets of \emph{locations} controlled by Player
    Min and Player Max and we write $L = L_\mMIN \cup L_\mMAX$; 
    $\act$ is a finite set of \emph{actions};
    $\clocks$ is a finite set of \emph{clocks}; 
    $\inv : L \to \zones$ is an \emph{invariant condition}; 
    $E : L {\times} \act \to \zones$ is an \emph{action enabledness function};
    $\rho : \act \to 2^{C}$ is a \emph{clock reset function}; 
    $\delta : L {\times} \act \to L$ is a  \emph{transition function}; and
    $p: L \cup L {\times} \act \to \Real$ is a \emph{price information function}. 
A PTGA is binary-priced when $p(\ell) \in \set{0, 1}$ for all
$\ell \in L$.
\end{definition}
When we consider a PTGA as an input of an algorithm, its size is understood as
the sum of the sizes of encodings of $L$, $\clocks$, $\inv$, $\act$, $E$, $\rho$,
$\delta$ and $p$.
We draw the states of $\mMIN$ players as circles, while states of
$\mMAX$ player  as boxes.

Let $\pta = (L_\mMIN, L_\mMAX, \act, \clocks, \inv,  E, \rho, \delta, p)$
be a PTGA.
A \emph{configuration} of a PTGA is a
pair $(\ell, \nu)$, where $\ell$ is a location and $\nu$ a clock valuation such that 
$\nu \sat \inv(\ell)$. 
For any $t \in \Rplus$, we let $(\ell,\nu){+}t$ equal the configuration
$(\ell,\nu{+}t)$.   
In a configuration $(\ell,\nu)$, a timed action (time-action pair)
$(t,a)$ is available if and only if the invariant condition $\inv(\ell)$ 
is continuously satisfied while $t$ time units elapse, and $a$ is enabled
(i.e.\ the enabling condition $E(\ell,a)$ is satisfied) after $t$ time units
have elapsed.  
Furthermore, if the timed action $(t,a)$ is performed, then the next
configuration is determined by the transition relation $\delta$ and the reset
function $\rho$,
i.e. the clocks in $\rho(a)$ are reset and we move to the location
$\delta(\ell, a)$. 

A game on a PTGA starts in an {\em initial configuration} $(\ell,\nu)\in L\times 
V$ and players $\mMin$ and $\mMax$ construct an infinite play by taking turns to
choose available  timed actions $(t, a)$ whenever the current location is controlled by
them and the price $p(\ell) \cdot t + p(\ell, a)$ is paid to the Max by
player Min.
Formally, PTGA semantics is given as a game arena. 

\begin{definition}[PTGA Semantics]\label{ptgsem-def}
  Let $\pta = (L_\mMIN, L_\mMAX, \act, \clocks, \inv,  E, \rho, \delta, p)$
  be a PTGA.   
  The semantics of $\pta$ is given by game arena 
  $\sem{\pta} {=} (S, S_\mMIN, S_\mMAX,  A, T, \pi)$ where 
  \begin{itemize}
  \item 
    $S \subseteq L {\times} V$ is the set of states such
    that $(\ell,\nu) \in S$ if and only if $\nu \sat \inv(\ell)$;
  \item
    $(\ell,\nu) \in S_\mMIN$ (or $(\ell,\nu) \in S_\mMAX$) 
    if $(\ell, \nu) \in S$ and $\ell \in L_\mMIN$ (or  $\ell\in
    L_\mMAX$, respectively). 
    \item 
    $A = \act {\times} \Rplus$ is the set of \emph{timed actions};
  \item
    $T : S \times \act \to S$ is the transition function
    such that for $(\ell, \nu) \in S$ and $(a, t) \in \act$, we
    have $T( (\ell, \nu)  , (a, t) ) = (\ell', \nu')$ if and
    only if 
    \begin{itemize}
    \item
      $\nu {+} t' \in Inv(\ell)$ for all $t' \in [0, t]$;
         $\nu {+} t \in E(\ell,a)$;
      $(\ell', \nu') \in S$,  
      $\delta(\ell, a) = \ell'$, $(\nu+t)_{\rho(a)} = \nu'$.
    \end{itemize}
  \item 
    $\pi : S{\times} \act {\to} \Real$ is the reward function where
    $\pi( (\ell, \nu), (a, t)) {=} p(\ell, a) + p(\ell) \cdot t$.
  \end{itemize}
\end{definition}

We are interested in the mean-payoff decision problem for timed automata $\pta$
that asks to decide whether the value of the mean-payoff game for a given state
is below a given budget. 
For a PTGA $\pta$ and budget $r \in \Real$, we write $\MPG(\pta, r)$ for the
$r$-mean payoff decision problem that asks whether value of the game at the
state $(\ell, \bf{0})$ is smaller than $r$. 
The following theorem summarizes the key contribution of this paper. 
\begin{theorem}
  The decision problem $\MPG(\pta, r)$ for binary-priced timed
  automata $\pta$ is undecidable for  automata with three clocks, and decidable for
  automata with one clock. 
\end{theorem}

\section{Boundary Region Graph Abstraction}
\label{sec:brg}
In this section we introduce an abstraction of priced timed games called the
boundary region abstraction (that generalizes classical corner-point
abstraction~\cite{BBL04}), and characterize conditions under which a solution of
optimality equations for the boundary region abstraction can be lifted to a 
solution of optimality equations for timed automata.
Observe that in order to keep our result as general as possible, we present the
abstraction and corresponding results for timed automata with an arbitrary number
of clocks.
In the following section, we show that the required conditions hold for the case
of one-clock binary-priced timed automata. 

\noindent {\bf Timed Successor Regions.}
Recall that $\Rr$ is the set of clock regions.
For $\region, \region' \in \Rr$, we say that $\region'$ is in the future of
$\region$, denoted $\region \xrightarrow{*} \region'$, 
if there exist $\nu \in \region$, $\nu' \in \region'$ and $t \in \Rplus$ such
that $\nu' = \nu{+}t$ and say $\region'$ is the \emph{time successor} of
$\region$ if $\nu{+}t' \in \region \cup \region'$ for all $t' \leq t$  and write 
$\region \rightarrow \region'$, or equivalently $\region' \leftarrow \region$,
to denote this fact.
For regions $\region, \region' \in \Rr$ such that $\region \xrightarrow{*}
\region'$ we write $[\region, \region']$ for the zone $\bigcup \{ \region''
  \, | \,  \region \xrightarrow{*} \region'' \wedge \region''
  \xrightarrow{*} \region' \}$.

\noindent {\bf Thin and Thick Regions.}
We say that a region $\region$ is \emph{thin} if 
$[\nu] {\neq} [\nu{+}\varepsilon]$ for every $\nu \in \region$ and 
$\varepsilon {>} 0$ and \emph{thick} otherwise.
We write $\Rr_\THIN$ and $\Rr_\THICK$ for the sets of thin and thick
regions, respectively. 
Observe that if $\region \in \Rr_\THICK$ then, for any $\nu \in \region$, there
exists $\varepsilon {>} 0$, such that $[\nu] {=} [\nu {+} \varepsilon]$ and the
time successor of a thin region is thick, and vice versa. 


\noindent{\bf Intuition for the Boundary Region Graph (BRG).}  Recall that $K$ is an upper bound on
clock values and let $\NATS{K} = \set{0, 1, \dots, K}$. 
For any $\nu \in V$, $b \in \NATS{K}$ and $c \in \clocks$, we define
$\ttime(\nu, (b,c)) {\rmdef} b {-} \nu(c)$ if $\nu(c) {\leq} b$, and
$\ttime(\nu, (b,c)) {\rmdef} 0$ if $\nu(c) {>} b$.
Intuitively, $\ttime(\nu, (b, c))$ returns the amount of time that must
elapse in $\nu$ before the clock $c$ reaches the integer value $b$. 
Observe that, for any $\region' \in \Rr_\THIN$, there exists $b \in \NATS{K}$
and $c \in \clocks$, such that $\nu \in \region$ implies  $(\nu {+} (b {-} \nu(c)) \in \region'$ for all $\region \in \Rr$ in the past of
$\region'$ and write $\region \rightarrow_{b, c} \region'$.
The boundary region abstraction is motivated by the following. 
Consider $a \in \act$, $(\ell,\nu)$ and $\region \xrightarrow{*}
\region'$ such that $\nu \in \region$, $[\region, \region'] \subseteq
\inv(\ell)$ and $\nu' \sat E(\ell, a)$. (For illustration, see Figure~\ref{fig:finboundary} in
Appendix).
\begin{itemize}
\item
  If $\region' \in \Rr_\THICK$,  then there are infinitely many $t \in \Rplus$
  such that  $\nu {+}t \in \region'$.
  However, amongst all such $t$'s, for one of the boundaries of $\region'$, the
  closer $\nu {+} t$ is to this boundary, the `better' the timed action 
  $(t, a)$ becomes for a player's objective.
  However, since $\region'$ is a thick region, the set $\set{t \in \Rplus \, |
    \, \nu {+} t \in \region'}$  is an open interval, and hence does not
  contain its boundary values. 
  Let the closest boundary of $\region'$ from $\nu$
  be defined by the hyperplane $c = b_{\sinf}$ and the farthest boundary
  of  $\region'$ from $\nu$ be defined by the hyperplane $c= b_{\ssup}$.
$b_{\sinf}, b_{\ssup} \in \mathbb{N}$ are such that 
$b_{\sinf} - \nu(c)$ ($b_{\ssup} {-} \nu(c)$) is the infimum (supremum) of the time spent to reach 
the lower (upper) boundary of region $\zeta'$. 
  Let the zones that correspond to these boundaries be denoted by $\region'_{\sinf}$
  and $\region'_{\ssup}$ respectively.
  Then $\region \to_{b_{\sinf}, c} \region'_{\sinf} \rightarrow \region'$
 and  $\region \to_{b_{\ssup}, c} \region'_{\ssup} \leftarrow \region'$.
  In the boundary region abstraction we include these `best'  timed
  actions through $(b_{\sinf}, c, a, \region')$ and 
  $(b_{\ssup},c, a, \region')$. 
\item
  If $\region' \in \Rr_\THIN$, then there exists a unique $t \in \Rplus$ such
  that $\nu{+}t \in \region'$. 
  Moreover since $\region'$ is a thin region, there exists a clock $c \in C$
  and a number $b \in \Nat$ such that $\region \to_{b, c} \region'$ and 
  $t = b {-} \nu(c)$.
  In the boundary region abstraction we summarise this `best' timed action from
  region $\region$ via region $\region'$ through the action 
  $(b, c, a, \region')$. 
\end{itemize}
Based on this intuition above the boundary region abstraction (BRA) is defined as
follows. 
\begin{definition}\label{brg-def}
  For a priced timed game arena
  $\pta = (L_\mMIN, L_\mMAX, \act, \clocks, \inv,  E, \rho, \delta, p)$
  the boundary region abstraction of $\pta$ is given by
  the game arena
  $\RegB{\pta} = (\hS, \hS_\mMIN, \hS_\mMAX, \hA, \hT, \hpi)$
   \begin{itemize}
    \item
    $\hS \subseteq L {\times} V {\times} \Rr$ is the set of
    states such that $(\ell, \nu, \region) \in \hS$ if and only if 
    $\region \subseteq \inv(\ell)$ and $\nu \in \CLOS{\region}$ (recall that
    $\CLOS{\region}$ denotes the closure of $\region$); 
  \item
    $(\ell,\nu, \region) \in \hS_\mMIN$ (or $(\ell,\nu, \region) \in \hS_\mMAX$) 
    if $(\ell, \nu, \region) \in \hS$ and $\ell \in L_\mMIN$ (or  $\ell\in
    L_\mMAX$, resp.). 
  \item
    $\hA = (\NATS{K} {\times} \clocks {\times} \act {\times} \Rr)$ is the set of
    actions;
  \item 
    For $\hs {=} (\ell, \nu, \region) {\in} \hS$ and $\alpha {=} (b_\alpha, c_\alpha, a_\alpha,
    \region_\alpha) {\in} \hA$, function $\hT(\hs, \alpha)$ is
    defined if  $[\region, \region_\alpha] {\subseteq} \inv(\ell)$ and
      ${\region_\alpha \subseteq E(\ell, a_\alpha)}$ and it equals
    $(\ell',\nu',\region') \in \hS$ where
    $\delta(\ell, a_\alpha) = \ell'$,
    $\nu_\alpha[C{:=}0]=\nu'$ and $\zeta_\alpha[C{:=}0]=\region'$ with
    $\nu_\alpha = \nu {+} \ttime(\nu, (b_\alpha,c_\alpha))$   
    and one of
    the following conditions holds: \\
      $\region \rightarrow_{b_\alpha, c_\alpha} \zeta_\alpha$; 
      $\region \rightarrow_{b_\alpha, c_\alpha} \region_{\sinf} \rightarrow \zeta_\alpha$ for
      some $\region_{\sinf} \in \Rr$; 
      $\region \rightarrow_{b_\alpha, c_\alpha} \region_{\ssup} \leftarrow \zeta_\alpha$ for
      some $\region_{\ssup} \in \Rr$;
  \item
    for  $(\ell, \nu, \region) \in \hS$ and $(b_\alpha, c_\alpha, a_\alpha,
    \region_\alpha) \in \hA$  the reward function $\hpi$ is given by: \\
   $ \hpi((\ell, \nu, \region),(b_\alpha, c_\alpha, a_\alpha,
    \region_\alpha)) = p(\ell, a_\alpha) + p(\ell)
    \cdot (b_\alpha {-} \nu(c_\alpha))$
  \end{itemize}
\end{definition}

Although the boundary region abstraction is not a finite game arena, every state
has only finitely many time-successor (the boundaries of the regions) 
and for a fixed initial state we can restrict attention to a finite game arena
due to the following observation.
\begin{lemma}[~\cite{Tri09}.]
  \label{prop:reachable-subgraph-is-finite-game}
  Let $\pta$ be a priced timed game arena and $\RegB{\pta}$ the corresponding BRA. 
  For any state of $\RegB{\pta}$, its reachable sub-graph is
  finite and can be constructed in time exponential in the size of $\pta$
  when $\pta$ has more than one clock.
  For one clock $\pta$, the reachable sub-graph of $\RegB{\pta}$
  can be constructed in time polynomial in the size of $\pta$.
  Moreover, the reachable sub-graph from the initial location and clock valuation is
  precisely the corner-point abstraction. 
\end{lemma}

\subsection{Reduction to Boundary Region Abstraction}
\label{brg-gb}
In what follows, unless specified otherwise, we fix a
PTGA $\pta = (L_\mMIN, L_\mMAX, \act, \clocks, \inv,  E, \rho, \delta, p)$
with semantics 
$\sem{\pta} {=} (S, S_\mMIN, S_\mMAX,  A, T, \pi)$
and BRA 
$  \RegB{\pta}  = (\hS, \hS_\mMIN, \hS_\mMAX, \hA, \hT, \hpi)$.
Let $G: \hS \to \Real$ and $B: \hS \to \Real$ be  such that
$(G, B) \models \Opt(\RegB{\pta})$, i.e. 
for every $\hs \in \hS$ we have that 
\begin{eqnarray*}
  G(\hs) & = & 
  \begin{cases}
    \max_{\alpha \in \hA(\hs)}  \set{G(\hs') \::\: \hs \xrightarrow{\alpha} \hs' } &  \text{ if $\hs \in \hS_\mMAX$}\\
    \min_{\alpha \in \hA(\hs)} \set{G(\hs') \::\: \hs \xrightarrow{\alpha} \hs' } &   \text{ if $\hs \in \hS_\mMIN$}.
  \end{cases}\\
  B(\hs) & = & 
  \begin{cases}
    \max_{\alpha \in \hA(\hs)}  \set{ \pi(\hs, \alpha) - G(\hs) + B(\hs') \::\: \hs \xrightarrow{\alpha} \hs'
      \text{ and } G(\hs) = G(\hs')} &  \text{ if $\hs \in \hS_\mMAX$}\\
    \min_{\alpha \in \hA(\hs)} \set{\pi(\hs, \alpha) - G(\hs) + B(\hs') \: : \: \hs \xrightarrow{\alpha} \hs'
      \text{ and } G(\hs) = G(\hs')} &   \text{ if $\hs \in \hS_\mMIN$}.
  \end{cases}
\end{eqnarray*}
For a function $F: \hS \to \Real$ we define a function $F^\boxplus:S \to
\Real$ as $(\ell, \nu) \mapsto  F(\ell, \nu, [\nu])$.
In this section we show under what conditions we can lift a solution $(G, B)$ of
optimality equations of BRA to $(\Gp, \Bp)$ for priced timed game arena.
%
Given a set of valuations $X {\subseteq} V$, a function $f : X \to \Rplus$ is
\emph{affine} if for any valuations $\nu_x, \nu_y \in X$ we have that for all
$\lambda \in [0, 1]$,
$f(\lambda \nu_x + (1-\lambda) \nu_y) =
\lambda f(\nu_x) + (1-\lambda) f(\nu_y)$.
We say that a function $f : \hS \to \Rplus$ is regionally affine if $f(\ell,
\cdot,\region)$ is affine over a region for all $\ell  \in L$ and  $\region \in \Rr$,
and $f$ is regionally constant if $f(\ell, \cdot,\region)$ is constant over a
region for all $\ell  \in L$ and $\region \in \Rr$. 
Some properties of affine functions that are useful in the proof of the key
lemma  are given in Lemma \ref{lem-keyprop}. 

\begin{lemma} \label{lem-keyprop}
\label{lem:prop}
  Let $X \subseteq V$ and $Y \subseteq \Rplus$ be convex sets.
  Let $f: X \to \Real$ and $w: X \times Y \to \Real$ be
  affine functions. Then for $C \subseteq \clocks$ we have
  that $\phi_C(\nu, t) = w(\nu, t) + f((\nu+t)[C{:=}0])$ is also an affine function, and
   $ \inf_{t_1 < t < t_2} \phi_C(\nu, t) = \min \set{\overline{\phi}_C(\nu, t_1),
      \overline{\phi}_C(\nu, t_2)} ~\text{and }~$\\
    $\sup_{t_1 < t < t_2} \phi_C(\nu, t) = \max \set{\overline{\phi}_C(\nu, t_1),
      \overline{\phi}_C(\nu, t_2)},$
     $\overline{\phi}$ is the unique continuous closure of $\phi$. 
\end{lemma}

\begin{theorem}
\label{thm:the-key-thm}
  Let $G: \hS \to \Real$ and $B: \hS \to \Real$ are such that
$(G, B) \models \Opt(\RegB{\pta})$ and $G$ is regionally constant and $B$ is
regionally affine, then $(\Gp, \Bp) \models \Opt(\pta)$.
\end{theorem}
\begin{proof}
  We need to show that $(\Gp, \Bp) \models \Opt(\pta)$, i.e. for every
  \begin{eqnarray*}
    \Gp(s) & = & 
     \begin{cases}
      \sup\limits_{(t, a) \in A(s)}  \set{\Gp(s') \::\: s \xrightarrow{(t, a)} s' } &  \text{ if $s \in S_\mMAX$}\\
      \inf\limits_{(t, a) \in A(s)} \set{\Gp(s') \::\: s \xrightarrow{(t, a)} s' } &   \text{ if $s \in S_\mMIN$}.
    \end{cases}\\
    \Bp(s) & = & 
    \begin{cases}
      \sup\limits_{(t, a) \in A(s)}  \set{ \pi(s, (t, a)) {-} \Gp(s) {+} \Bp(s') \::\: s
        \xrightarrow{(t, a)} s'
        \text{ and } \Gp(s) = \Gp(s')} &  \text{ if $s \in S_\mMAX$}\\
      \inf\limits_{(t, a) \in A(s)} \set{\pi(s, (t, a)) {-} \Gp(s) {+} \Bp(s') \: : \: s
        \xrightarrow{(t, a)} s'
        \text{ and } \Gp(s) = \Gp(s')} &   \text{ if $s \in S_\mMIN$}.
    \end{cases}
   \end{eqnarray*}
   Consider the case when $s = (\ell, \nu) \in S_\mMIN$ and consider
the right side  of the gain equations. 
\begin{eqnarray*}
  && \inf_{(t, a) \in A(s)} \set{\Gp(s') \::\: s \xrightarrow{(t, a)} s' } \\
  & = & \min_{\substack{\region'': [\nu] \rightarrow^{*} \region'' \\ [\region,
        \region''] \in \inv(\ell)}}
  \min_{a \in \act} \inf_{\substack{t \::\:\\ \nu+t \in \region''}}
  \set{G(\delta(\ell, a), (\nu{+}t)[\rho(a){:=}0], [(\nu{+}t)][\rho(a){:=}0]) }\\
   &=&  \min_{\alpha \in \hA(\ell, \nu, [\nu])} \set{G(\ell', \nu', \region')
     \::\: (\ell, \nu, \region) \xrightarrow{\alpha} (\ell', \nu',\region')}
  = G(\ell, \nu, [\nu]) = \Gp(\ell, \nu).
\end{eqnarray*}
The first equality holds since $(G, B) \models \Opt(\RegB{\pta})$.
The second equality follows since $G$ is regionally constant and hence it
suffices to consider the delay $\ttime(\nu, (b, c))$ that corresponds to either
left or right boundary of the region $\region''$, 
i.e. for fixed $\nu,
\region''$ and $a \in \act$ we have that 
$\inf_{\substack{t \::\:\\ \nu+t \in \region''}}
\set{G(\ell', (\nu+t)[\rho(a){:=}0], \region')}
= G(\ell', \nu_\alpha[C{:=}0], \region') $
where  $\nu_\alpha = \nu {+} \ttime(\nu, (b_\alpha,c_\alpha))$, $\zeta''[C{:=}0]=\region'$ with
$\region \rightarrow_{b_\alpha, c_\alpha} \zeta''$ if $\zeta''$ is thin, 
and   $\region \rightarrow_{b_\alpha, c_\alpha} \region_{\sinf} \rightarrow \zeta''$ for
some $\region_{\sinf} \in \Rr$ if $\zeta''$ is thick.
Similarly, for the bias equations, we need to show:
\begin{eqnarray*}
&&\inf_{\substack{t \::\:\\ \nu+t \in \region''}}
\set{\pi((\ell, \nu), (t, a)) - G(\ell, \nu) + B (\ell', (\nu+t)[\rho(a){:=}0], \region')}\\
&=& \pi((\ell, \nu, [\nu]), (\ttime(\nu, (b_\alpha,c_\alpha)))) - G(\ell, \nu, [\nu]) + B(\ell', \nu_\alpha[C{:=}0], \region') 
\end{eqnarray*}
where  $\nu_\alpha = \nu {+} \ttime(\nu, (b_\alpha,c_\alpha))$, $\zeta''[C{:=}0]=\region'$ with
$\region \rightarrow_{b_\alpha, c_\alpha} \zeta''$ if $\zeta''$ is thin;
and   $\region \rightarrow_{b_\alpha, c_\alpha} \region_{\sinf} \rightarrow
\zeta''$ for some $\region_{\sinf} \in \Rr$ 
or $\region \rightarrow_{b_\alpha, c_\alpha} \region_{\ssup} \rightarrow \zeta''$
for some $\region_{\ssup} \in \Rr$ if $\zeta''$ is thick.
Given $B$ is regionally affine (and hence linear in $t$) and the price function
is linear in $t$, the whole expression
$\pi((\ell, \nu), (t, a)) - G(\ell, \nu) + B (\ell',
(\nu+t)[\rho(a){:=}0], \region')$ is linear in $t$ and from Lemma~\ref{lem:prop}
it  attains its  infimum or supremum  on either boundary of the region.
\end{proof}

\section{Decidability for One Clock Binary-priced PTGA}
\label{sec:dec}
Given the undecidability with 3 or more clocks, we focus on one clock PTGA.
We provide a strategy improvement algorithm to compute a solution
$G: \hS \to \Real$ and $B: \hS \to \Real$ of the optimality equations, i.e. 
$(G, B) \models \Opt(\RegB{\pta})$ for the BRA  $\RegB{\pta}  = (\hS, \hS_\mMIN,
\hS_\mMAX, \hA, \hT, \hpi)$ of one-clock binary-priced PTGAs with certain
``integral payoff'' restriction. 
Further, we show that for one clock binary-priced integral-payoff PTGA, 
the solution of optimality equations of corresponding BRG is such that the gains
are regionally  constant and biases are regionally affine. 
Hence by Theorem \ref{thm:the-key-thm}, the algorithm can be applied to solve
mean-payoff games for one-clock binary-priced integral-payoff PTGAs.
We also show how to lift the integral-payoff restriction to recover decidability
for one-clock binary-priced PTGA. 

\noindent{\bf Regionally constant positional strategies.}
Standard strategy improvement algorithms iterate over a finite set of strategies
such that the value of the subgame at each iteration gets strictly improved. 
However, since there are infinitely many positional strategies in a boundary region
abstraction, we focus on ``regionally constant'' positional strategies (\RCPSs).
We say that a positional strategy $\mu: \hS \to \hA$ of player Min is
regionally-constant  if for all $(\ell, \nu, \region), (\ell, \nu', \region)
\in \hS_\mMIN$ we have that $[\nu] {=} [\nu']$ implies that $\mu(\ell, \nu, \region)
= \mu(\ell, \nu',\region)$.
We similarly define \RCPSs{}  for player Max.
In other words, in an \RCPS{} a player chooses same boundary action for every
valuation of a region---as a side-result we show that optimal strategies for
both players have this form.  
Observe that there are finitely many \RCPSs{} for both players.
We write $\hPi_\mMIN$ and $\hPi_\mMAX$ for the set of \RCPSs{} for player Min and
player Max, respectively. 
For a BRA  $\hpta$, $\chi \in \hPi_\mMAX$, and $\mu \in \hPi_\mMIN$  we write
$\hpta(\chi)$  and $\hpta(\mu)$ for the ``one-player'' game on the sub-graph of
BRAs  where the strategies of player Max and Min have
been fixed to \RCPSs{} $\chi$ and $\mu$, respectively. 
Similarly we define the zero-player game $\hpta(\mu, \chi)$ where strategies of
both players are fixed to \RCPSs{} $\mu$ and $\chi$.

Let $\hpta(\chi, \mu)$ be a zero-player game on the subgraph
where strategies of player Max (and Min) is fixed to \RCPSs{}  $\chi$ (and $\mu$).
Observe that for $\hpta(\mu, \chi)$ the unique runs originating from states
$\hs_0 = (\ell, \nu, \region)$ and $\hs'_0 = (\ell, \nu', \region)$ with $[\nu]
= [\nu']$ follow the same ``lasso'' after one step, i.e. the unique runs 
$\hs_0 \xrightarrow{\alpha_1}  \hs_1 \cdots
  \hs_k (\xrightarrow{\alpha_{k+1}} \cdots \hs_{k+N-1} \xrightarrow{\alpha_{k+N}}
  \hs_k)^*$  
and 
$\hs'_0 \xrightarrow{\alpha_1}  \hs'_1  \cdots
  \hs'_k (\xrightarrow{\alpha_{k+1}} \cdots \hs'_{k+N-1} \xrightarrow{\alpha_{k+N}}
  \hs'_k)^*$   
are such that for $\hs_i = (\ell_i, \nu_i, \region_i)$ and $\hs_i' =
(\ell_i', \nu_i', \region_i')$ we have that  $\ell_i =
\ell_i'$, $\region_i = \region_i'$ and $\nu_i = \nu_i'$ for all $i \in [1,k{+}N{-}1]$.
This is so because for one-clock timed automata the successors of the states
$\hs_0 = (\ell, \nu, \region)$ and $\hs_0'=(\ell, \nu', \region)$ for  action
$\alpha_1 =(b, c, a, \region')$ is the same $(\ell'', \nu'', \region'')$ where
$\nu''(c) = \nu(c) + (b - \nu(c)) = b = \nu(c) + (b - \nu(c))$ if $c \not \in
\rho(a)$ and $\nu''(c) = 0$ otherwise.    
Consider the optimality equations (Section~\ref{app:oe}) for the lasso. 
%
Observe that the gain for the states $\hs_0, \ldots, \hs_{k+N-1}$ is the same,
and let's call it $g$.
If we add the bias equations side-wise for the cycle, we get
$g = \frac{1}{N} \sum_{i=0}^{N-1} \pi(\hs_{k+i}, \alpha_{k+i+1})$. 
It follows from the previous observation that the gains are regionally
constant.

\noindent{\bf Integral Payoff PTGA.}
The gain in a zero-player game, $\hpta(\chi, \mu)$, although
regionally-constant,  may not be a whole number. 
We say that a PTGA is integral-payoff if for every pair
$(\mu, \chi) \in \hPi_\mMIN \times \hPi_\mMAX$ of \RCPSs{} the gain as defined
above is a whole number.
Observe that the denominator in the gains correspond to the number of edges in a simple
cycle of the BRA $\RegB{\pta}$. 
If there are $N$ simple cycles in the region graph of length $n_1, n_2 \dots,
n_N$, then let $\mathcal{L}$ be the least-common multiple  of $n_1, n_2 \dots, n_N$.
We multiply constants appearing in the guards and invariants of the timed
automata by $\mathcal{L}$.
It is easy to observe that mean-payoff of any state in the original PTGA  $\pta$
is the mean-payoff in $\Upsilon_\pta$ divided by $\mathcal{L}$.  
For notational convenience, we assume that the given PTGA is an integral-payoff
PTGA and hence for \RCPS{} strategy profile $(\mu, \chi)$ the gain is
regionally constant and integral. 


\begin{algorithm}[t]
  Consider $\hpta(\mu, \chi)$ as a (single successor) weighted graph $\Gg = (V, E, w)$ where 
  \begin{itemize}
  \item $V = L \times \Rr \times \Rr$ (with an order $\preceq$)  and $E \subseteq V \times \hA \times V$
  \item
    $(v_1, \alpha, v_2) \in E$ if $v_1 = (\ell_1,  \region_1, \region_1')$, $v_2=(\ell_2,
    \region_2,\region'_2)$, and   $\mu(\ell_1, \nu_1, \region_1') = \alpha$ (or $\chi(\ell_1, \nu_1, \region_1')
    = \alpha$) 
    for all $\nu_1 \in \region_1$ and
    $(\ell_1, \nu_1, \region_1') \xrightarrow{\alpha} (\ell_1, \nu_2, \region_2')$
    for some $\nu_2 \in\region_2$.
  \item
    $w(v_1, \alpha, v_2)$ is the expression $\nu \mapsto b_\alpha - \nu(c_\alpha)$\;
  \end{itemize} 
  \For{every cycle $C$ of $\Gg$}
  {
    Let $\REACH(C)$ be set of vertices that reach $C$\;
    Let $\gamma$ be the average weight of the cycle ($w$ is constant on cycles)\;
    For every vertex $V$ in $\REACH(C)$ set $G(V) = \gamma$ and $B(V) = \bot$\;
    For the smallest $\preceq$-vertex $V_*$ in $C$. Set $B(V_*) = 0$\; 
       \While{there is $V \in \REACH(C)$ with $B(V) = \bot$}
          {
            Let $(V', \alpha, V'') \in E$ with $B(V'') \not = \bot$\;
            $B(V') := \nu \mapsto \left(w(V', \alpha, V'')(\nu) - G + B(V'')\right)$\; 
          }
   } 
    \Return $(G, B)$\;
  \caption{\textsc{ComputeValueZeroPlayer}$(\pta, \mu, \chi)$}
  \label{alg2}                                     
\end{algorithm}
\subsection{Strategy Improvement Algorithm for Binary-Priced PTGA} \label{subsec-algo}
Let $\pta$ be a one-clock integral-payoff binary-priced PTGA $\pta$ and $\hpta$
be its boundary region graph.
For a given \RCPS{} profile $(\mu, \chi) \in \hPi_\mMIN \times \hPi_\mMAX$,
Algorithm~\ref{alg2} computes the solution for the optimality equations
$\Opt(\pta(\mu, \chi))$.
This algorithm considers $\hpta(\mu, \chi)$ as a graph whose vertices are
``regions'' $(\ell, [\nu], \region)$ corresponding to state
$(\ell, \nu, \region) \in \hS$  of the boundary region graph, edges are boundary
actions between them determined by the regionally constant strategy profile, and
weight of an edge is the time function associated with the boundary action.
Observe that every cycle in this graph will have constant weight on the edges
since taking boundary actions in a loop will require going from an integral
valuation to another integral valuation, and the average cost of such a cycle
can be easily computed. 

Also observe that, not unlike standard convention~\cite{Put94}, our algorithm
chooses a vertex in a 
cycle  arbitrarily and fixes the bias of all of the states in that vertex to $0$.
This is possible since optimality equations over a cycle are underdetermined,
and we exploit this flexibility to achieve  solution to biases in a particularly
``simple'' structure. 
We say that a function $f : \hS \to \Rplus$ is regionally simple~\cite{AM99}  if
for all $\ell \in L$, $\region, \region' \in \Rr$ either i) there exists a
$d\in \Nat$ such that $f(\ell, \nu, \region') = d$ for all $\nu \in \region$;
or ii) there exists $d \in \Nat$ and $c \in \clocks$ such that
$f(\ell, \nu, \region') = d - \nu(c)$ for all $\nu \in \region$.
Key properties of regionally simple functions (Lemma~\ref{lem:simple} in
Appendix \ref{app-lem_simple}) include that they are also regionally affine,
closed under minimum and maximum,  and if $B: \hS \to \Real$ be a regionally
simple function and $G: \hs \to \Nat$ be a regionally constant function,
then $\hs \mapsto \pi(\hs, \alpha) - G(\hs) + B(\hs')$ where
$\hs \xrightarrow{\alpha} \hs'$ is a regionally simple function.  
Using these properties and induction on the distance to $\preceq$-minimal
element in the reachable cycle, we prove the correctness and following property
of Algorithm~\ref{alg2}. 
\begin{lemma}
  \label{lem:zero-player}
  Algorithm~\ref{alg2} computes solution of optimality equations
  $(G, B) \models \Opt(\hpta(\mu, \chi))$ for $\mu \in \hPi_\mMIN$ and
  $\chi \in \hPi_\mMAX$. Moreover, $G$ is regionally constant and $B$
  is regionally simple. 
\end{lemma}

\begin{algorithm}[t]
  Choose an arbitrary regionally constant positional strategy
  $\chi' \in \Pi_\mMAX$\;
  \Repeat{$\chi = \chi'$}{
    $\chi := \chi'$\;
    Choose an arbitrary regionally constant positional strategy
    $\mu' \in \Pi_\mMIN$\;
    \Repeat{$\mu = \mu'$}{
      $\mu := \mu'$\;
      $(G, B) := \textsc{ComputeValueZeroPlayer}(\pta, \mu, \chi)$ \;
      $\mu' := \textsc{ImproveMinStrategy}(\pta, \mu, G, B)$ \; 
    }
    $\chi' := \textsc{ImproveMaxStrategy}(\pta, \chi, G, B)$ \;
  }
  \Return $(G, B)$\;
    \vspace{-0.3em}
  \caption{\textsc{ComputeValueTwoPlayer}$(\pta)$}
  \label{SIA}
\end{algorithm}

\noindent The strategy improvement algorithm to solve optimality equations is given as
Algorithm~\ref{SIA}.
It begins by choosing an arbitrary regionally constant positional strategy
$\chi'$ and at every iteration of the loop ($2$--$11$) the algorithm computes
($5$--$9$) the value $(G, B)$ of the current \RCPS{} $\chi$ and based on the value,
the function $\textsc{ImproveMaxStrategy}$ returns an improved strategy by
picking boundary action that lexicographically maximizes gain and bias
respecting the policy that switches a decision only for a strict improvement. 
We formally define the function $\textsc{ImproveMaxStrategy}$ as follows: 
for $\chi {\in} \hSigma_\mMAX$, $G: \hS {\to} \Real$, and $B: \hS {\to} \Real$ we
let strategy $\textsc{ImproveMaxStrategy}(\pta, \chi, G, B)$ be such
that for all $\hs \in \hS_\mMAX$ we have 
\begin{eqnarray*}
  \textsc{ImproveMaxStrategy}(\pta, \chi, G, B)(\hs)
  =
  \begin{cases}
    \chi(\hs) & \text{ if $\chi(\hs) \in M^*(\hs, G, B) $} \\
    \text{Choose}(M^*(\hs, G, B)) & \text{ Otherwise}. 
  \end{cases}
\end{eqnarray*}
where $ M^*(\hs, G, B) =  \argmaxlex_{\alpha \in \hA} \set{(G(\hs'), \pi(\hs,
  \alpha) - G(\hs) + B(\hs')) \::\: \hs \xrightarrow{\alpha} \hs'}$ and
$\text{Choose}$ picks an arbitrary element from a set. 
\textsc{ImproveMaxStrategy} satisfies the following.
\begin{lemma}
  \label{lemma2}
  If $\chi \in \hPi_\mMAX$, $G$ is regionally constant, and $B$ is regionally
  simple, then function  $\textsc{ImproveMaxStrategy}(\pta, \chi, G, B)$ returns a regionally constant
  positional strategy.
\end{lemma}



The lines ($5$--$9$) compute the value of the strategy $\chi$ of Player Max via 
a strategy improvement algorithm.
This sub-algorithm works by starting with an arbitrary strategy of Player Min
and computing the value $(G, B)$ of the zero-player PTGA $\hpta(\mu, \chi)$.
Based on the value, the function $\textsc{ImproveMinStrategy}$
returns an improved strategy of Min.
The function $\textsc{ImproveMinStrategy}$ is defined as a dual of the function
$\textsc{ImproveMaxStrategy}$  where $\chi$ is replaced by $\mu$ and $\argmax$
by $\argmin$. 
\textsc{ImproveMinStrategy} satisfies the following.
\begin{lemma}
  \label{lemma1}
  If $\mu \in \hPi_\mMIN$, $G$ is regionally constant, and $B$ is regionally
  simple, then function  $\textsc{ImproveMinStrategy}(\pta, \chi, G, B)$ returns
  a regionally constant positional strategy.
\end{lemma}
It follows from Lemma~\ref{lemma2} and Lemma~\ref{lemma1} that at every iteration of the
strategy improvement the strategies $\mu$ and $\chi$ are \RCPSs{}. 
Together with finiteness of the set of \RCPSs{} and strict improvement at every
step (Lemma~\ref{impr1} and~\ref{impr2}), we get  following result.
\begin{theorem}
\label{thm:final}
  Algorithm~\ref{SIA} computes solution of optimality equations
  $(G, B) \models \Opt(\hpta)$ for integral payoff PTGA $\pta$.
  Moreover, $G$ is regionally constant and $B$ is regionally affine.  
\end{theorem}
This theorem---together with Theorem~\ref{thm:the-key-thm} and
Theorem~\ref{thm:gain-bias-correctness}---gives a proof of decidability for
mean-payoff games for integral-payoff binary-priced one-clock timed automata.

\section{Undecidability Results}
\label{sec:undec}
\begin{theorem}
  \label{thm_undec}
The mean-payoff problem $\MPG(\pta, r)$ is undecidable for PTGA
$\pta$ with 3 clocks having location-wise  price-rates $\pi(\ell) \in \{0,1,-1\}$ for
all $\ell \in L$ and $r = 0$.
Moreover, it is undecidable for binary-priced $\pta$ with 3 clocks and $r >= 0$.
\end{theorem}
\begin{proof}
  We first show the undecidability result of the mean-payoff problem $\MPG(\pta, 0)$
  with location prices $\{1,0,-1\}$ and no edge prices.
  We prove the result by reducing the non-halting problem of  2 counter
  machines.
  Our reduction uses a PTGA  with 3 clocks $x_1, x_2, x_3$,  location prices
  $\{1,0,-1\}$, and no edge prices.
  Each counter machine instruction (increment, decrement, zero check) is
  specified using a  PTGA module.
  The main invariant in our reduction is that on entry into any module, 
  we have $x_1 =   \frac{1}{5^{c_1}7^{c_2}}$, $x_2 = 0$ and $x_3=0$, where $c_1,c_2$ are
  the values of counters $C_1,C_2$.
  We outline the construction for the decrement instruction of counter $C_1$
  in Figure \ref{fig_undec_mpg_dec_new}.
  For conciseness, we present here modules using arbitrary location prices. However, 
  we can redraw these with extra locations and edges using only the location prices 
  from $\{1,0,-1\}$ as shown for $WD_1^1$ in Figure~\ref{redraw} in Appendix. 
 
  The role of the Min player  is to   faithfully simulate the two counter machine, by choosing appropriate
  delays to adjust the clocks to reflect changes in counter values.
  Player Max will have the opportunity to verify that player $\mMIN$ did not
  cheat while simulating the machine.  

  \begin{figure}[t]
  \centering
  \begin{tikzpicture}[->,>=stealth',shorten >=1pt,auto,node distance=1cm,
      semithick,scale=0.8]
  \node[initial,initial text={}, player1] at (0,-.5) (lk) {$0$ } ;
   \node()[above of=lk,node distance=5mm]{$\ell_k$};

  \node[player2] at (3,-.5) (chk){$0$} ;
     \node()[above of=chk,node distance=5mm]{$\text{Check}$};

  \node () [below right of=chk,node distance=6.5mm,xshift=-3mm] {$[x_3=0]$};

   \node[player1] at (6,-.5)(lk1){$\ell_{k+1}$};

\node[widget] at (-2,-.5) (W1){$WD^1_1$};

\node[widget] at (8,-.5) (W2){$WD^1_2$};

  \node[initial, initial text={},player1] at (-3,-2.5)(A) {$-5$};
   \node()[above of=A,node distance=5mm]{$A$};

  \node[player1] at (-0.5,-2.5) (B){$20$};
   \node()[above of=B,node distance=5mm]{$B$};

  \node[player1] at (2,-2.5) (C){$-15$};
   \node()[above of=C,node distance=5mm]{$C$};

  \node[player1] at (4.5,-2.5) (D) {$0$};
   \node()[above of=D,node distance=5mm]{$D$};

  \node[player1] at (7,-2.5) (E) {$0$};
   \node()[above of=E,node distance=5mm]{$E$};

  \path (lk) edge node {$x_1 {\leq} 1$} node[below] {$\set{x_3}$}(chk);
  \path (chk) edge node[below] {$\set{x_2}$} (lk1);
  \path (chk) edge[bend left] node {} (W1);
  \path (chk) edge[bend left] node {} (W2);

  \path (A) edge node {$ x_2 {=} 1$} node[below]{$\set{x_2}$} (B);
  \path (B) edge node {$x_1 {=} 2$} node[below]{$\set{x_1}$} (C);
  \path (C) edge node {$x_1 {=} 1$} node[below]{$\set{x_1}$} (D);
  \path (D) edge node {$x_2 {=} 2$} node[below]{$\set{x_2}$} (E);
  \draw[rounded corners] (E) -- (7, -3.8) --
  node[above] {$x_3=3, \{x_3\}$} (-3, -3.8) --  (A);

  \node[rotate=90] at (-4.5, -2.5) (N) {$\text{WD}^1_1$};
  \draw[dashed,draw=gold,rounded corners=10pt] (-4,-4.2) rectangle (7.5,-1.6);
  
 \end{tikzpicture}
\caption{Simulation to decrement counter $C_1$, mean cost is $\varepsilon$ for
  error $\varepsilon$. The widget $\text{WD}^1_2$ has exactly the same structure
  and guards on all transitions as $\text{WD}^1_1$, but the price signs  are reversed.}
\vspace{-1em}
\label{fig_undec_mpg_dec_new}
\end{figure}
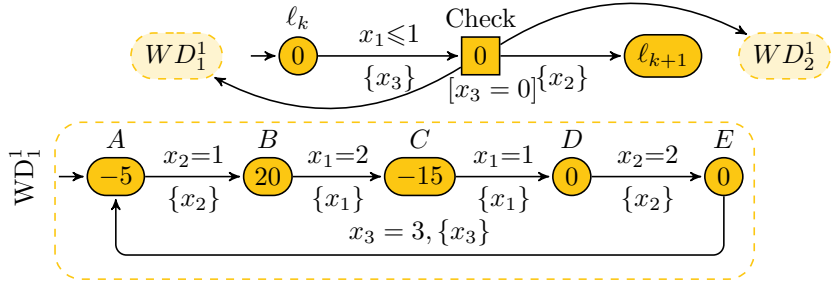

  We enter location $\ell_k$ with
$x_1=\frac{1}{5^{c_1}7^{c_2}}, x_2 = 0$ and $x_3=0$. Lets denote by
$x_{old}$ the value $\frac{1}{5^{c_1}7^{c_2}}$.  To correctly
decrement $C_1$, player Min should choose a delay of $4x_{old}$ at
location $\ell_k$.  At location $\text{Check}$, there is no time
elapse and player Max has three possibilities : ($i$) to go to
$\ell_{k+1}$ and continue the simulation, or ($ii$) to enter the widget $\text{WD}^1_1$, or (iii) to enter the widget $\text{WD}^1_2$.  If
player Min makes an error, and delays $4x_{old}+\varepsilon$ 
or $4x_{old}-\varepsilon$ at
$\ell_k$ ($\varepsilon > 0$), then player Max can enter one of the widgets
and punish player Min. Player Max enters widget $\text{WD}^1_1$ if 
the error made by player Min is of the form 
$4x_{old}+\varepsilon$ at $\ell_k$ 
and   enters widget $\text{WD}^1_2$ if the error made by player Min 
is of the form $4x_{old}-\varepsilon$ at
$\ell_k$. 

Let us examine the widget $\text{WD}^1_1$.
When we enter $\text{WD}^1_1$ for
the first time, we have $x_1=x_{old}+4x_{old}+\varepsilon$,
$x_2=4x_{old}+\varepsilon$ and $x_3=0$.  In $\text{WD}^1_1$, the cost
of going once from location $A$ to $E$ is $5\varepsilon$. Also, when we
get back to $A$ after going through the loop once, the clock values
with which we entered $\text{WD}^1_1$ are restored; thus, each time,
we come back to $A$, we restore the starting values with which we
enter $\text{WD}^1_1$.  The third clock is really useful for this
purpose only. It can be seen that the mean cost of 
transiting from $A$ to $A$ through $E$ is $\varepsilon$.  In a similar way, it can be checked 
that the mean cost of transiting from $A$ to $A$ through $E$
in widget $\text{WD}^1_2$ is $\varepsilon$ when player Min chooses a delay 
$4x_{old}-\varepsilon$ at $\ell_k$. 
Thus, if player Min makes a simulation error, player 
Max can always choose to goto one of the widgets, and ensure that the mean pay-off is not $\leq 0$.  
Note that when $\varepsilon=0$, then player Min will achieve his
objective: the mean pay-off will be 0.
Details of other gadgets are in Appendix~\ref{app:undec}. 
\end{proof}

In the Appendix~\ref{app:undec_per_time_unit}, we show how this undecidability results
extends (with the same parameters) if one defines mean payoff per time unit
instead of per step.  
This way of averaging across time spent was considered in~\cite{BCR14}, where 
the authors show the undecidability of $\MPG(\pta, 0)$ with 5 clocks.
We improve this result to show undecidability already in $3$ clocks. 




\newpage
\appendix
\centerline{\Large Appendix}
\section{Supplementary material to Section~\ref{sec:prelims}}

\subsection{Strategy improvement algorithm for Finite Game Arenas}
Let $\Gamma$ be a  finite game arena.
For technical convenience let us fix an arbitrary but fixed linear order
${\preceq} \subseteq S^2$ on the set of states~$S$.
For a positional strategy $\chi \in \Pi_\mMAX$ we write $\Gamma_\chi$ for the
subgame of $\Gamma$ where the outgoing transitions from the states controlled by
Player $\mMAX$ have been restricted to the ones allowed by $\chi$. 
We similarly define $\Gamma_\mu$ and $\Gamma{\chi \mu}$. 

\noindent{\bf Strategy Improvement Algorithm for Finite Game Arenas.} The
strategy improvement algorithm to compute s solution of optimality equations 
works as follows.
\begin{enumerate}
\item
 Fix an arbitrary positional strategy $\chi : S_\mMAX \to A$ for player~Max. 
\item
 (\emph{Best counter-strategy against $\chi$}.) 
 Compute the best counter-strategy $\mu$ for player~Min against the 
 strategy $\chi$ by performing the following steps.
 \begin{enumerate}
 \item
   (\emph{Minimize gain}.)
   For every state $s \in S$, let $G(s)$ be the value of the minimum
   average weight of a cycle reachable from the state $s$ in the
   strategy subgame $\Gamma_\chi$; 
   let $C(s)$ denote such a cycle reachable from the state~$s$.
   Let $m_s$ be the $\preceq$-smallest state on the cycle~$C(s)$.
   Let $W_s$ be the set of states $s'$, such that $m_{s'} = m_s$, i.e.,
   the set of states which have the same reachable minimum average
   weight cycle in the graph $\Gamma_\chi$ as the state $s$.
 \item
   (\emph{Minimize bias}.)
   For every state $s \in S$, let $B(s)$ be the weight of the 
   shortest
   path from the state $s$ to the state $m_s$ in the
   subgraph of $\Gamma_\chi$ induced by the set of states $W_s$, and
   where $G(s)$ was subtracted from the price of every transition in the
   subgraph $\Gamma_\chi$.
   For every state $s \in S_\mMIN$, set $\mu(s)$ to be the state  
   $s' \in V$, such that $s \xrightarrow{a}  s'$ for some $a$ , $s' \in W_s$ 
   (and hence $G(s) = G(s')$), and $B(s) = (\pi(s, a) - G(s)) + B(s')$. 
 \end{enumerate}
 Observe that the functions $G$ and $B$ thus obtained satisfy the
 optimality equations for the subgame $\Gamma_\chi$. 
 However, these function may not satisfy the optimality equations for the
 original game $\Gamma$. 
 The next step ``locally'' changes the strategy~$\chi$,
 intuitively in order to make progress towards computing the optimality
 equations. 
 
\item
 (\emph{Local improvement of strategy $\chi$}.)
 For every state $s \in S_\mMAX$, set $\chi(s)$ to be a successor
 $s'$ of the state~$s$, which first maximizes $G(s')$ and then 
 maximizes~$B(s')$.   
 In other words, $\chi(s)$ is the successor $s'$ of the state $s$
 which maximizes $(G(s'), B(s'))$ according to the lexicographical
 ordering on pairs, where we use the usual ordering on the reals in
 both coordinates.
 Importantly, if the current $\chi$-successor of the state $s$ is
 already maximum in the above lexicographic sense, $\chi(s)$
 remains unchanged, even if there are other successors $s'$ of $s$
 with the same pair of values $(G(s'), B(s'))$ as the
 state~$\chi(s)$.  
 \emph{This assumption is important for finite termination of the strategy
   improvement algorithm.}
 
\item
 If the local improvement of the strategy $\chi$ in the previous
 step resulted in a change of $\chi$ in at least one state then
 go back to step~2.
 Otherwise stop.
\end{enumerate}

We establish the following two fundamental properties of the
iterative scheme of strategy improvement described above.
The first observation that a locally optimal strategy yields a
solution to optimality equations is straightfoward to check.

\begin{lemma}[OE Solution from a locally optimal
   strategy] 
 \label{proposition:witness-from-termination}
 If the algorithm stops then the tuple $(G, B)$
 computed in the last iteration is a solution to optimality equations. 
\end{lemma}
Next, we show that in every non-terminating iteration of the algorithm,
the pair $(G, B)$ consisting of the gain function $G$ and the bias 
function $B$, that are uniquely determined from the current pair of
strategies $\chi$ and $\mu$, strictly increases according to a
certain linear ordering as a result of the local improvement.
This implies finite termination of the algorithm, since there are 
only finitely many positional strategies.
Thus, together with
Lemma~\ref{proposition:witness-from-termination}, we get the
existence of an OE Solution, which establishes
positional determinacy of mean-payoff games on finite game arenas.  

\begin{theorem}[Strict global improvement from myopic improvement]
\label{lemma:global-improvement}
 Let $\mu$ ($\mu'$) be the best counter-strategy for player~Min
 against a strategy $\chi$ ($\chi'$) for player~Max, and let $G$
 ($G'$) and $B$ ($B'$) be as computed in step~2 of an
 iteration of the algorithm starting from the strategy $\chi$
 ($\chi'$.) 
 If the strategy $\chi'$ is a non-trivial local improvement of the
 strategy $\chi$, as computed in step~3 of the algorithm, then for
 every state $s \in S$, the following hold.
 \begin{enumerate}
 \item
   We have $G'(s) \geq G(s)$.
 \item
   If $G'(s) = G(s)$ then $B'(s) \geq B(s)$.
 \item
   If $s {\in} S_\mMAX$ and $\chi'(s) {\not=} \chi(s)$ 
   then either  
   $G'(s) {>} G(s)$, or $G'(s) {=} G(s)$ and $B'(s) {>} B(s)$.
 \end{enumerate}  
\end{theorem}
\begin{proof}
  In order to verify property~1 it suffices to show that the average
  weight of every cycle reachable from a state $s$ in the strategy
  subgraph $\Gamma_{\chi'}$ is no smaller than the smallest average weight
  of a cycle reachable from the state $s$ in the strategy subgraph
  $\Gamma_{\chi}$.
  First, observe that for every transition $(s, a, s')$ in the subgraph of the  
  graph $\Gamma_{\chi'}$, we have the inequality $G(s) \leq G(s')$.
  It implies in $\Gamma_{\chi'}$ we have that $G(s)$ is smaller
  than average of the cheapest reachable cycle.
  On the other hand $G'(s)$ is the average of the cheapest cycle in
  $\Gamma_{\chi'}$. It follows that $G(s) \leq G(s')$. 

  Now we argue that the properties~2 and~3 hold.
  From the assumption that $G'(s) = G(s)$ it follows that the paths from
  the state $s$ in graphs $\Gamma{\chi \mu}$ and $\Gamma_{\chi' \mu'}$
  lead to the same cycle.
  We need to prove that $B'(s) \geq B(s)$ and that 
  $\chi'(s) \not= \chi(s)$ implies $B'(s) > B(s)$.
  
  First, observe that for every transition $(s, a, s')$ in the subgraph of the  
  graph $\Gamma_\chi$ induced by the set of states  $W_s$, we have the
  inequality $B(s) \leq (\pi(s, a) - G(s)) + B(s')$;
  it follows by the construction of $B$ as the weights of shortest
  paths to the state $s$ (step 2(b)).
  Moreover, by the definition of the myopic improvement of the strategy
  $\chi$ (step~3), for every transition $(s, a, s')$ in the subgraph of the
  graph $\Gamma_{\chi'}$ induced by the set of states $W'_s$, we have
  $B(s) \leq (\pi(s, a) - G(s)) + B(s')$.
  
  Now let $\seq{s = s_0, a_1, s_1, \dots, s_{p} = m_v}$ be a path from the
  state $s$ to the state $m_s$ in the subgraph of the graph
  $\Gamma_{\chi'}$ induced by the set of states $W'_s$.
  Then adding the $p$ inequalities 
  $B(s_i) \leq (\pi(s_i, a_{i+1}) - G(s_i)) + B(s_{i+1})$, for $i = 0, 1, \dots,
  p-1$, we get that $B(s) \leq \sum_{i=0}^{p-1} (\pi(s_i, a_{i+1}) - G(s))$.
  This, however, implies that $B(s) \leq B'(s)$, since if the previous
  inequality holds for \emph{all} the paths from $s$ to $m_s$ in the
  appropriate subgraph of $\Gamma_{\chi'}$, then it also holds for the 
  shortest such. 
  This establishes property~2.
  Property~3 now follows from the strictness of the inequality 
  $B(s_0) < (\pi(s_0, a_1) - G(s_0)) + B(s_1)$ if we assume that  
  $\chi'(s) \not= \chi(s)$.
  Note that this is when the assumption at the end of step~3 is
  necessary to avoid looping without strict improvement of neither the
  gain nor the bias function from one iteration of the algorithm to
  another. 
\end{proof}

\section{Boundary Region Abstraction: Illustration}
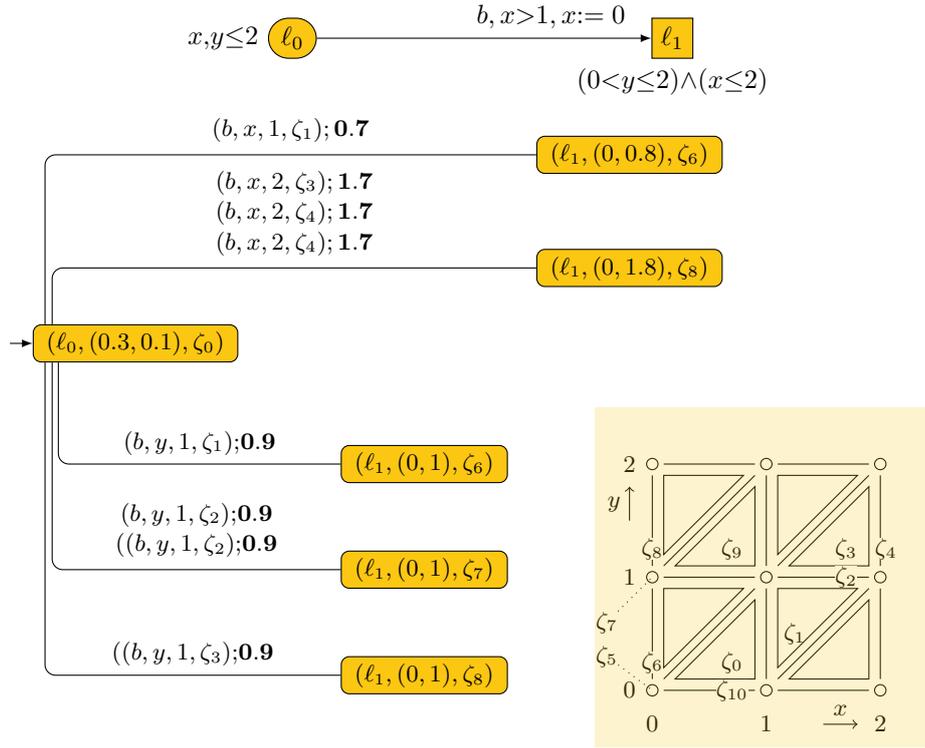
\begin{figure}[t]
  \centering
\begin{tikzpicture}[node distance=2cm,auto,->,>=latex]
  \node[player1, label={left:$x{,}y{\le}2$}] at (0,0) (l0)  {$\ell_0$};
  \node[player2, label={below:$(0{<}y{\le}2){\wedge}(x{\le}2)$}] at (5,0) (l1) {$\ell_1$};
  
  \path  (l0) edge  node [pos=0.7, above] {$b, x{>}1, x{:=0}$} (l1);

\end{tikzpicture}
\begin{tikzpicture}
\tikzstyle{every node}=[font=\small]
\tikzstyle{loc}=[rounded corners=3pt,fill=gold,draw,minimum size=1.4em,inner sep=0em]
\tikzstyle{trans}=[-latex, rounded corners]

\node[loc] at (0,0) (l0) {$\begin{array}{c}(\ell_0, (0.3,0.1),\region_0)\end{array}$};
\node[loc] at (6.5,2.5) (l1) {$\begin{array}{c}(\ell_1, (0,0.8), \region_6)\end{array}$};
\node[loc] at (6.5,1) (l2) {$\begin{array}{c}(\ell_1, (0,1.8), \region_8)\end{array}$};
\node[loc] at (3.8,-1.6) (l3) {$\begin{array}{c}(\ell_1, (0,1), \region_6)\end{array}$};
\node[loc] at (3.8,-3) (l4) {$\begin{array}{l}(\ell_1, (0,1), \region_7)\end{array}$};
\node[loc] at (3.8,-4.4) (l5) {$\begin{array}{l}(\ell_1, (0,1),\region_8 )\end{array}$};

 \draw[trans] ($(l0.-180) - (0.3,0)$) -- (l0);
 \draw[rounded corners] (l0.-192)|-(l1) node[pos=0.75,above]
  {$\begin{array}{c}(b, x, 1, \region_1);\mathbf{0.7}\end{array}$};    
 \draw[rounded corners] (l0.-193)|-(l2) node[pos=0.75,above]
      {$\begin{array}{c}(b, x, 2, \region_3);\mathbf{1.7}\\(b, x, 2,
          \region_4);\mathbf{1.7}\\(b, x, 2, \region_4);\mathbf{1.7}\end{array}$};
 \draw[rounded corners] (l0.194)|-(l3) node[pos=.75,above] {$(b, y, 1, \region_1){;}\mathbf{0.9}$};
 \draw[rounded corners] (l0.193)|-(l4) node[pos=.75,above] {$\begin{array}{c}(b,
     y, 1, \region_2){;}\mathbf{0.9}\\((b, y, 1, \region_2){;}\mathbf{0.9}\end{array}$};
 \draw[rounded corners] (l0.192)|-(l5) node[pos=.75,above] {$((b, y, 1, \region_3){;}\mathbf{0.9}$};

\node (X) at (8.3,-3.1) {
\begin{tikzpicture}[x=1.5cm,y=1.5cm]
\foreach \i in {0,1}
{
 \foreach \j in {0,1}
 {
   \draw (\i+0.1,\j+0) -- (\i+0.9,\j+0);
   \draw (\i+0,\j+0.1) -- (\i+0,\j+0.9);
   \draw (\i,\j) circle (0.05);
   \draw (\i+0.2,\j+0.1) -- (\i+0.9,\j+0.1) -- (\i + 0.9, \j + 0.8) -- cycle;
   \draw (\i+0.1,\j+0.2) -- (\i+0.8,\j+0.9) -- (\i + 0.1, \j + 0.9) -- cycle;
   \draw (\i+0.1,\j+0.1) -- (\i+0.9,\j+0.9);
 }  
}
\foreach \i in {0,1}
{
 \foreach \j in {2}
 {
   \draw (\i+0.1,\j+0) -- (\i+0.9,\j+0);
   \draw (\i,\j) circle (0.05);
 }  
}
\foreach \i in {2}
{
 \foreach \j in {0,1}
 {
   \draw (\i+0,\j+0.1) -- (\i+0,\j+0.9);
   \draw (\i,\j) circle (0.05);
 }  
}
\draw (2,2) circle (0.05);

 \node[font=\small] at (0.7,0.25) {$\region_0$};
 \node[font=\small] at (1.25,0.5) {$\region_1$};
 \node[font=\small,fill=white,inner sep=0] at (1.7,1) {$\region_2$};
 \node[font=\small] at (1.7,1.25) {$\region_3$};
 \node[font=\small,fill=white,inner sep=0] at (2.05,1.25) {$\region_4$};
 \node[font=\small,fill=white,inner sep=0] (r5) at (-0.4,0.3) {$\region_5$};
 \draw[dotted] (r5) -- (0,0);
 \node[font=\small,fill=white,inner sep=0] at (0,0.25) {$\region_6$};
 \node[font=\small,fill=white,inner sep=0] (r7) at (-0.4,0.6) {$\region_7$};
 \draw[dotted] (r7) -- (0,1);
 \node[font=\small,fill=white,inner sep=0] at (0,1.25) {$\region_8$};
 \node[font=\small] at (0.7,1.25) {$\region_9$};
 \node[font=\small,fill=white,inner sep=0] at (0.7,0) {$\region_{10}$};

 \node at (-0.2,0) {$0$};
 \node at (-0,-0.3) {$0$};
 \node at (-0.2,1) {$1$};
 \node at (-0.2,2) {$2$};
 \node at (1,-0.3) {$1$};
 \node at (2,-0.3) {$2$};
\draw[->] (1.5,-0.3) -- (1.8,-0.3) node[midway,above] {$x$};
\draw[->] (-0.2,1.5) -- (-0.2,1.8) node[midway,left] {$y$};

\fill[gold,fill opacity=0.2] (-0.5,-0.5) rectangle (2.5,2.5);

\end{tikzpicture}};
\end{tikzpicture}
    \caption{Sub-graph of the boundary region abstraction for the PTGA with the
      region names as depicted in the bottom right
      corner.}\label{fig:finboundary}   
\end{figure}

A PTGA is shown at the top of Figure~\ref{fig:finboundary}. 
A sub-graph of BRA reachable
from $(\ell_0, (0.3,0.1), 0{<}y {<} x{<}1)$ is shown below the PTGA in the same figure.
The names of the regions correspond to the regions depicted in the bottom right
corner.
Edges are labelled $(a, c, b, \region)$ and the intuitive meaning is to
wait until clock $c$ reaches the value $b$ in the boundary of the
region $\region$.
Considering the region $\region_1$, we see that it is determined by the
constraints $(1{<}x{<}2) {\wedge} (0{<}y{<}1)  {\wedge} (y{<}x{-}1)$.
The bold numbers on edges correspond to the time delay before the action
labelling the edge is taken.
Figure~\ref{fig:finboundary} includes the actions available in the
initial state and one of the action pairs that are available in the state
$(\ell_1, (0,1), (x{=}0){\wedge}(1{<}y{<}2))$.

\section{Proofs from Section~\ref{sec:dec}}
\subsection{Examples of PTGAs with non-affine Bias Functions}
\label{app:lem-bias}
\begin{example}
Consider the timed game shown in Figure~\ref{fig:bias}. 
\begin{figure}[t]
\centering
\begin{minipage}[b]{0.45\linewidth}
    \scalebox{1}{
      \begin{tikzpicture}[node distance=2cm,auto,->,>=latex] 
        \node[player1,initial,initial text={}](2){\makebox[0mm][c]{$1$}};%
        \node[player1](4)[below right of=2,xshift=8mm,yshift=.5cm]{\makebox[0mm][c]{$0$}};%
        \node[player1](5)[above right of=2,xshift=8mm,yshift=-.5cm]{\makebox[0mm][c]{$0$}};%
        \node[player1](7)[right of=4,xshift=10mm]{$1$};%
        \node[player1](8)[right of=5,xshift=10mm]{$1$};%

        \node()[above of=2,node distance=5mm,color=gray]{$\ell_1$};
        \node()[below of=7,node distance=5mm,color=gray]{$\ell_5$};
        \node()[above of=5,node distance=5mm,color=gray]{$\ell_2$};
        \node()[below of=4,node distance=5mm,color=gray]{$\ell_4$};
        \node()[above of=8,node distance=5mm,color=gray]{$\ell_3$};

        \path 
        (2) edge node[sloped,above,node distance=0mm,xshift=-1.2mm,yshift=1mm] {$1 < x < 2$} (5) %
        (2) edge node[sloped,below,yshift=-1mm, xshift=-1mm]{$x < 2$} (4)
        (4) edge[bend right=15] node[below,xshift=1mm]{$\{x\}$} (7)
        (7) edge[bend right=15] node[above,xshift=2mm]{$x=1, \{x\}$} (4)
        (5) edge[bend right=15] node[below,xshift=1mm]{$\{x\}$} (8)
        (8) edge[bend right=15] node[above,xshift=2mm]{$x=1, \{x\}$} (5);
      \end{tikzpicture}}
%
  \caption{One clock PTGA with binary prices may not have regionally simple bias.}
  \label{fig:bias}
\end{minipage}
\quad
\begin{minipage}[b]{0.46\linewidth}
    \scalebox{1}{
      \begin{tikzpicture}[node distance=2cm,auto,->,>=latex] 
        \node[player1,initial,initial text={}](2){\makebox[0mm][c]{$0$}};%
        \node[player2](4)[below right
        of=2,xshift=6mm,yshift=.5cm]{\makebox[0mm][c]{$1$}};%
        \node[player2](5)[above right
        of=2,xshift=6mm,yshift=-.5cm]{\makebox[0mm][c]{$1$}};%
        \node[player2](7)[below right
        of=5,xshift=6mm,yshift=.5cm]{$-1$};%

        \node()[above of=2,node distance=5mm,color=gray]{$\ell_1$};
        \node()[above of=7,node distance=5mm,color=gray]{$\ell_2$};
        \node()[above of=5,node distance=5mm,color=gray]{$\ell_3$};
        \node()[below of=4,node distance=5mm,color=gray]{$\ell_4$};

        \path 
        (2) edge node[above]{$x < 1$} (7)
        (5) edge[bend right=15] node[above,node
        distance=0mm,xshift=-1.2mm,yshift=1mm]
        {$x < 1,\{x\}$} (2) %
        (4) edge[bend left=15] node[below,yshift=-1mm, xshift=-1mm]{$x < 1$} (2)%
        (7) edge[bend left=15] node[below,xshift=1mm]{$x < 1$} (4) %
        (7) edge[bend right=15] node[above,xshift=2mm]{$x=1, \{x\}$} (5);%
      \end{tikzpicture}}
%
  \caption{One clock PTGA with non-binary prices.
  The $\varepsilon$-optimal strategies of $\mMIN$ and $\mMAX$
  players are not regionally positional boundary strategies and
  the value of the game is $\frac{1}{4}$.}
  \label{fig:CE-0+1-1}
\end{minipage}
\end{figure}
All the locations here belong to the player $\mMIN$.
There are two cycles in which the gains or the average weight of the cycles is $\frac{1}{2}$.
If corresponding to two or more strategies, the gains are the same,
then the strategy that minimizes the bias is chosen by the player $\mMIN$.
The bias $B(s)$ of a player $\mMIN$ state $s = (\ell, \nu)$ where
$\ell$ is a location of the priced timed game and
$\nu$ is a clock valuation is given by
$\inf\limits_{(t, a) \in A(s)} \set{\pi(s, (t, a)) {-} G(s) {+} B(s') \: : \: s
        \xrightarrow{(t, a)} s'
        \text{ and } G(s) = G(s')}$ if $s \in S_\mMIN$.
The $\inf$ is replaced by a $\sup$ for a player $\mMAX$ state.
$G(s)$ denotes the gain of state $s$.
Considering the bias of the states $(\ell_2, \nu)$ and $(\ell_5, \nu)$ to be 0,
the bias of a state $(\ell_1, \nu)$ turns out to be
$\frac{1}{2}-\nu_x$ when the edge $(\ell_1 \rightarrow \ell_2)$
is considered and the bias of $(\ell_1, \nu)$ is the constant function $-1$
when the edge $(\ell_1 \rightarrow \ell_4)$ is considered.
Here $\nu_x$ denotes the value of clock $x$.
For $x > \frac{3}{2}$, the value of $\frac{1}{2}-\nu_x$ is smaller than $-1$
and hence the edge $(\ell_1 \rightarrow \ell_4)$ is chosen
while the edge $(\ell_1 \rightarrow \ell_2)$ is chosen when $x \le \frac{3}{2}$.
Thus the bias is not regionally affine.
This can be attributed to the fact that the gain is not integral.
\end{example}

\begin{example}
Let us consider another example of one-clock priced timed games with three price
rates: $0$, $1$ and $-1$ shown in Figure~\ref{fig:CE-0+1-1}.
The value of the game is $\frac 1 4$ which is obtained by the following
sequence of moves by both players.
From the initial state $(\ell_1, x=0)$,
player $\mMIN$ waits at $\ell_1$ for $0.5$ time units
and then moves the token to location $\ell_2$.
player $\mMAX$ moves the 
token to $\ell_4$ immediately and subsequently the token
reaches $\ell_1$ with the value of clock $x=1-\delta$, where
$\delta$ is an infintesimal positive quantity.
The token is forwarded to $\ell_2$ now instantaneously
or after an infintesimal delay so that the value of clock $x$ is still less than 1.
player $\mMAX$ now forwards the token to $\ell_3$ when the value of clock $x$ becomes 1.
At location $\ell_3$,  $1 - \delta$ amount of time is elapsed.
In the next move of player $\mMAX$, the token
reaches the initial state $(\ell_1, x= 0)$.
The $\varepsilon$-optimal strategies of both $\mMIN$ and player $\mMAX$
are not regionally constant boundary strategies.
\end{example}

\begin{lemma} \label{lem-bias}
In one clock binary non-integral payoff PTGA the bias may not be regionally affine.
The same is true for for the BRA of the PTGA.
\end{lemma}
\proof
Consider the example in Figure \ref{fig:bias}.
There are two loops, one consisting of locations $\ell_2$ and $\ell_3$
and the other one with locations $\ell_4$ and $\ell_5$.
For the states $(\ell_2, \nu)$ and $(\ell_3, \nu)$
where $\nu$ is any valuation of clock $x$,
the gain $g$ equals $\frac{1}{2}$.
For the states $(\ell_4, \nu)$ and $(\ell_5, \nu)$
such that $\nu_x \le 1$ also have their gain $g$ equal to $\frac{1}{2}$.
Hence the corresponding PTGA is a non-integral payoff one.
Note that all the states in the non-integral payoff PTGA in Figure \ref{fig:bias} belong to
the player $\mMIN$.
Since the gains corresponding to both the loops are the same,
the optimal strategy of the player $\mMIN$ from a state $(\ell_1, \nu)$
is determined by the bias of a successor state.

We now show that the bias of a state $(\ell_1, \nu)$ may not be regionally affine.
For computing the bias in the loop consisting of the locations $\ell_2$ and $\ell_3$,
let us consider that $B(\ell_2, \nu) = 0$.
Thus considering the successor state of a state $(\ell_1, \nu)$
to be the state $(\ell_2, \nu+t)$, we have
\begin{eqnarray*}
B(\ell_1, \nu) &=& 
\inf_{t: 1 < \nu + t < 2} \; t - g + B(\ell_2,\nu+t) \\
&=& \inf_{t: 1-\nu < t < 2-\nu}\;   t - \frac{1}{2} + 0 
= 1 - \nu - \frac{1}{2} = \frac{1}{2} - \nu.
\end{eqnarray*}
In the loop consisting of locations $\ell_4$ and $\ell_5$, let $B(\ell_5, \nu) = 0$.
Now
\[
 B(\ell_4, \nu) = \inf_{t} \; t - g + B(\ell_5,0) = 0 - \frac{1}{2} + 0 = -
 \frac{1}{2}.
 \]
Thus considering the successor state of a state $(\ell_1, \nu)$
to be the state $(\ell_4, \nu+t)$, we have
\[
B(\ell_1, \nu) = \inf_{t:0\le t < 2} \; t - g + B(\ell_4,\nu+t) = 0 -
\frac{1}{2} -\frac{1}{2} = -1.
\]
Hence the bias at $(\ell_1, \nu)$ is given by $\min (\frac{1}{2} - \nu_x, -1)$.
This gives us that the edge $\ell_1 \to \ell_2$ is chosen when $x \le \frac{3}{2}$,
while the edge $\ell_1 \to \ell_4$ is chosen when $x > \frac{3}{2}$.
Hence the bias of location $\ell_1$ is not regionally affine in non-integral payoff PTGA.
It is easy to see that the lemma also holds for the BRA on the PTGA.
\qed

\subsection{Regionally Simple Functions: Properties}
\label{app-lem_simple}
Simple functions were introduced by Asarin and Maler~\cite{AM99} in the context
of reachability timed games for timed automata.
\begin{lemma}[Simple Functions and Their Properties]
  \label{lem:simple}
  A function $f : V \to \Rplus$ is simple if 
  \begin{itemize}
  \item
    either, there exists $d \in \Nat$ such that for all $\nu \in V$ we have $f(\nu) = d$;
  \item
    or, there exists $d \in \Nat$ and $c \in \clocks$ such that for all $\nu \in V$ we have $f(\nu) = d - \nu(c)$.
  \end{itemize}
  We say that a function $f : \hS \to \Rplus$ is regionally simple  if
  for all $\ell \in L$, $\region, \region' \in \Rr$ either i) there exists a
  $d\in \Nat$ such that $f(\ell, \nu, \region') = d$ for all $\nu \in \region$;
  or ii) there exists $d \in \Nat$ and $c \in \clocks$ such that
  $f(\ell, \nu, \region') = d - \nu(c)$ for all $\nu \in \region$.
The simple functions have the following properties. 
  \begin{enumerate}
  \item
    Every simple function is an affine function.
  \item
    Every regionally simple function is a regionally affine function.
  \item
    For $\ell \in L$, $\region, \region' \in \Rr$ let $D_{\ell,\region,\region'}
    = \set{(\ell, \nu, \region') \::\: [\nu] = \region}$.  
    For simple functions $f, g: D_{\ell,\region,\region'} \to \Real$ we have
    that
    \begin{itemize}
    \item
      $\min \set{f, g}$ is either $f$ or $g$.
    \item
      $\max \set{f, g}$ is either $f$ or $g$.
    \end{itemize}
  \item 
    Minimum and maximum of a finite set of regionally simple functions is
    regionally simple.
  \item
    Let $B: \hS \to \Real$ be a regionally simple function and $G: \hs \to
    \Nat$ be a regionally constant function.
    Let 
    $B_\alpha^\oplus: \hS \to \Real$ as $\hs \mapsto \pi(\hs, \alpha) - G(\hs) +
    B(\hs')$ where $\hs \xrightarrow{\alpha} \hs'$.
    We have that $B_\alpha^\oplus$ is a regionally simple function. 
\end{enumerate}
\end{lemma}   
\begin{proof}
    The proof is in many parts.
  \begin{enumerate}
  \item
    Trivial.
  \item
    Trivial.
  \item
    Let $f, g: D_{\ell, \region, \region'} \to \Real$. There are four cases to
    consider. 
    \begin{itemize}
    \item
      For all $\nu \in \region$ we have $f(\nu) = d_1$ and $g(\nu) = d_2$.
      If $d_1 < d_2$ the $\min \set{f, g} = f$ else $\min \set{f, g} = g$.
      Similarly if $d_1 < d_2$ then $\max \set{f, g} = g$ else $\min \set{f, g}
      = f$. 
    \item
      For all $\nu \in \region$ we have $f(\nu) = d_1$ and $g(\nu) = d_2
      -\nu(c_2)$.
      Notice that if for some valuation $d_1 < d_2 - \nu(c_2)$ then for all
      valuations $d_1 < d_2 - \nu'(c_2)$ since the difference between various
      valuations in a region is strictly less than $1$. 
      The lemma follows for this case from this observation. 
    \item
      For all $\nu \in \region$ we have $f(\nu) = d_1-\nu(c_1)$ and $g(\nu) =
      d_2$.
      This case is analogous to previous case and omitted. 
    \item
      For all $\nu \in \region$ we have $f(\nu) = d_1-\nu(c_1)$ and $g(\nu) =
      d_2-\nu(c_2)$.
      If for some region $d_1 -\nu(c_1) < d_2 - \nu(c_2)$ then for all
      valuations this inequality will follow since  difference between various
       values of clocks from different valuations in a region is strictly smaller
       than $1$. 
    \end{itemize}
    
  \item
    Similar to property~3.
  \item
    Let $B: \hS \to \Real$ be a regionally simple function and $G: \hs \to
    \Nat$ be a regionally constant function.
    Let 
    $B_\alpha^\oplus: \hS \to \Real$ as $\hs \mapsto \pi(\hs, \alpha) - G(\hs) +
    B(\hs')$ where $\hs \xrightarrow{\alpha} \hs'$.
    Let $\hs = (\ell, \nu, \region)$ and $\hs' = (\ell', \nu', \region')$.
    Let $g \in \Nat$ be such that $G(\ell, \nu, \region) = g$.
    There are eight cases to consider.
    \begin{itemize}
    \item
      $B(\ell', \cdot, \region')$ is $\nu \mapsto d$,
      $p(\ell) = 0$ and $c_\alpha \not \in \rho(a_\alpha)$.
      In this case $B(\ell, \cdot, \region) = p(\ell, a_\alpha) -
      g + d$ (an integer) is a simple function. 
    \item
      $B(\ell', \cdot, \region')$ is $\nu \mapsto d $,
      $p(\ell) = 0$ and $c_\alpha \in \rho(a_\alpha)$ 
      In this case $B(\ell, \cdot, \region) =  p(\ell, a_\alpha) -
      g + d$ (an integer) is a simple function. 
    \item
      $B(\ell', \cdot, \region')$ is $\nu \mapsto d $,
      $p(\ell) = 1$ and $c_\alpha \not \in \rho(a_\alpha)$ 
      In this case $B(\ell, \cdot, \region) = (b_\alpha - \nu(c_\alpha)) +
      p(\ell, a_\alpha) - g + d = K - \nu(c_\alpha)$ is a simple function.
    \item
      $B(\ell', \cdot, \region')$ is $\nu \mapsto d $,
      $p(\ell) = 1$ and $c_\alpha \in \rho(a_\alpha)$ 
      In this case $B(\ell, \cdot, \region) =
      (b_\alpha - \nu(c_\alpha)) + p(\ell, a_\alpha) - g + d = K -
      \nu(c_\alpha)$ is a simple function. 
    \item
      $B(\ell', \cdot, \region')$ is $\nu \mapsto d - \nu(c)$,
      $p(\ell) = 0$ and $c_\alpha \not \in \rho(a_\alpha)$ 
      In this case $B(\ell, \cdot, \region) =
      p(\ell, a_\alpha) - g + d - (\nu(c) + b_\alpha - \nu(c_\alpha)) =
      p(\ell, a_\alpha) - g + d - b_\alpha$ (an integer) is a simple function.
      We have $\nu(c) = \nu(c_\alpha)$ as there in only one clock. 
    \item
      $B(\ell', \cdot, \region')$ is $\nu \mapsto d - \nu(c)$,
      $p(\ell) = 0$ and $c_\alpha \in \rho(a_\alpha)$
      In this case $B(\ell, \cdot, \region) =
      p(\ell, a_\alpha) - g + d$ (an integer) is a simple function.
    \item
      $B(\ell', \cdot, \region')$ is $\nu \mapsto d - \nu(c)$,
      $p(\ell) = 1$ and $c_\alpha \not \in \rho(a_\alpha)$ 
      In this case $B(\ell, \cdot, \region) =
      (b_\alpha - \nu(c_\alpha)) + p(\ell, a_\alpha) - g + d - (\nu(c) +
      b_\alpha - \nu(c_\alpha)) =  (b_\alpha - \nu(c_\alpha)) + p(\ell,
      a_\alpha) - g + d - b_\alpha = K -\nu(c_\alpha)$ is a simple function.  
    \item
      $B(\ell', \cdot, \region')$ is $\nu \mapsto d - \nu(c)$,
      $p(\ell) = 1$ and $c_\alpha \in \rho(a_\alpha)$ 
      In this case $B(\ell, \cdot, \region) =
      (b_\alpha - \nu(c_\alpha)) + p(\ell, a_\alpha) - g + d = K -
      \nu(c_\alpha)$ is a simple function.
    \end{itemize}
  \end{enumerate}
The proof is now complete.
\end{proof}

\subsection{Proof of Lemma~\ref{lem:zero-player}}
\label{app:oe}
Let $\hpta(\chi, \mu)$ be a zero-player game on the subgraph
where strategies of player Max (and Min) is fixed to \RCPSs{}  $\chi$ (and $\mu$).
We sketch an algorithm $\textsc{ComputeValueZeroPlayer}(\pta, \mu, \chi)$ which
returns the solution of optimality equations $\Opt(\pta(\mu, \chi))$.
Observe that for $\pta(\mu, \chi)$ the unique runs originating from states
$\hs_0 = (\ell, \nu, \region)$ and $\hs'_0 = (\ell, \nu', \region)$ with $[\nu]
= [\nu']$ follow the same ``lasso'' after one step, i.e. the unique runs 
\begin{eqnarray*}
\hs_0 &\xrightarrow{\alpha_1}&  \hs_1 \cdots
  \hs_k \left(\xrightarrow{\alpha_{k+1}} \cdots \hs_{k+N-1} \xrightarrow{\alpha_{k+N}}
  \hs_k\right)^*  
\text{ and } \\
  \hs'_0 &\xrightarrow{\alpha_1}&  \hs'_1  \cdots
  \hs'_k \left(\xrightarrow{\alpha_{k+1}} \cdots \hs'_{k+N-1} \xrightarrow{\alpha_{k+N}}
  \hs'_k\right)^*   
\end{eqnarray*}
are such that for $\hs_i = (\ell_i, \nu_i, \region_i)$ and $\hs_i' =
(\ell_i', \nu_i', \region_i')$ we have that  $\ell_i =
\ell_i'$, $\region_i = \region_i'$ and $\nu_i = \nu_i'$ for all $i \in [1,
  k+N-1]$.
This is so because for one-clock timed automata the successors of the states
$\hs_0 = (\ell, \nu, \region)$ and $\hs_0'=(\ell, \nu', \region)$ for  action
$\alpha_1 =(b, c, a, \region')$ is the same $(\ell'', \nu'', \region'')$ where
$\nu''(c) = \nu(c) + (b - \nu(c)) = b = \nu(c) + (b - \nu(c))$ if $c \not \in
\rho(a)$ and $\nu''(c) = 0$ otherwise.    
Consider the optimality equations for the lasso. 
\begin{eqnarray}
  G( \hs_0) &=& G(\hs_1) = \cdots = G(\hs_k) = \cdots G(\hs_{k+N-1}) = g \text{
    say }\nonumber\\
  B( \hs_0) &=& \pi(\hs_0, \alpha_1) - g + B(\hs_1) \nonumber  \\
  B( \hs_1) &=& \pi(\hs_1, \alpha_2) - g + B(\hs_2) \nonumber \\
  \dots && \dots \label{eqis} \\
  B( \hs_k) &=& \pi(\hs_k, \alpha_{k+1}) - g + B(\hs_{k+1}) \nonumber \\
  B( \hs_{k+1}) &=& \pi(\hs_{k+1}, \alpha_{k+1}) - g + B(\hs_{k+3}) \nonumber\\
  \dots && \dots \nonumber\\
  B( \hs_{k+N-1}) &=& \pi(\hs_{k+N-1}, \alpha_{k+N}) - g + B(\hs_k) \nonumber
\end{eqnarray}

  We show this via an induction on the distance from the region whose biases we
  fixed earlier to $0$.
  Let $\hs_0 \xrightarrow{\alpha_1}  \hs_1 \xrightarrow{\alpha_2} \cdots
  \hs_k$ be the run such that $\hs_i = (\ell_i, \nu_i, \region_i)$ and for all
  states $B(\ell_k, \cdot, \region_k) = 0$.
  We prove via induction that $B(\ell_i, \cdot, \region_i)$ is a simple function.
  The base case follows easily since $B(\ell_k, \cdot, \region_k) = 0$ which is a
  simple function.
  Assume that $B(\ell_i, \cdot, \region_i)$ is simple we show that
  $B(\ell_{i-1}, \cdot, \region_{i-1})$ defined as
  \[
  \nu \mapsto p(\ell_{i-1}) \cdot (b_\alpha - \nu(c_\alpha)) + p(\ell_{i-1},
  a_\alpha) - g + B(\ell_i, \nu_\alpha, \region_i)
  \]
  is also simple where $\alpha_i = (b_\alpha, c_\alpha, a_\alpha,
  \region_\alpha)$.
  This follows form property~5 of Lemma~\ref{lem:simple}.

\subsection{Proof of Lemma~\ref{lemma2} and Lemma~\ref{lemma1}}
\begin{proof}
  To prove this lemma it suffices to show that the function 
  \[
  M^*(\cdot, G, B): \hs \mapsto  \argmaxlex_{\alpha \in \hA} \set{(G(\hs'), \pi(\hs,
    \alpha) - G(\hs) + B(\hs')) \::\: \hs \xrightarrow{\alpha} \hs'}
  \]
  is regionally constant.
  We already have that $G$ is regionally constant.
  From the property~5 of Lemma~\ref{lem:simple} it follows that $\hs \mapsto
  \pi(\hs, \alpha) - G(\hs) + B(\hs'))$ is regionally simple.
  Note that since maximum (Property 3 of Lemma~\ref{lem:simple} of any finite
  set of simple functions over a region is one of the functions from the set, it
  follows that the set $M^*(\cdot, G, B)$ is regionally constant. 
  The proof for the Lemma~\ref{lemma1} is similar and hence omitted. 
\end{proof}

\subsection{Proof of Theorem~\ref{thm:final}}
The following two Lemmas together with finiteness of regionally constant positional
strategies give the proof of Theorem~\ref{thm:final}.
The proofs of these lemma are similar to Lemma~\ref{lemma:global-improvement}
and hence omitted. 
\begin{lemma}
  \label{impr1}
  Let $(G, B) = \Opt(\pta(\chi))$. 
  If $\chi' = \textsc{ImproveMaxStrategy}(\pta, \chi, G, B)$ then $(G' B') =
  \Opt(\pta(\chi'))$ is such that $(G', B') \geq (G, B)$ and if $\chi' \not =
  \chi$ then $(G' , B') > (G, B)$. 
\end{lemma}
\begin{lemma}
    \label{impr2}
  Let $(G, B) = \Opt(\pta(\chi, \mu))$. 
  If $\mu' = \textsc{ImproveMinStrategy}(\pta, \mu, G, B)$ then $(G' B') =
  \Opt(\pta(\chi, \mu'))$ is such that $(G', B') \leq (G, B)$ and if $\mu' \not =
  \mu$ then $(G' , B') < (G, B)$. 
\end{lemma}

\section{Proofs from Section~\ref{sec:undec}}
We prove our undecidability result using a reduction from two-counter machine. 
A two-counter machine $M$ is a tuple $(L, C)$ where ${L = \set{\ell_0,
    \ell_1, \ldots, \ell_n}}$ is the set of instructions---including a
distinguished terminal instruction $\ell_n$ called HALT---and ${C =
  \set{c_1, c_2}}$ is the set of two \emph{counters}.  The
instructions $L$ are one of the following types:
\begin{enumerate}
\item (increment $c$) $\ell_i : c := c+1$;  goto  $\ell_k$,
\item (decrement $c$) $\ell_i : c := c-1$;  goto  $\ell_k$,
\item (zero-check $c$) $\ell_i$ : if $(c >0)$ then goto $\ell_k$
  else goto $\ell_m$,
\item (Halt) $\ell_n:$ HALT.
\end{enumerate}
where $c \in C$, $\ell_i, \ell_k, \ell_m \in L$.
A configuration of a two-counter machine is a tuple $(l, c, d)$ where
$l \in L$ is an instruction, and $c, d$ are natural numbers that specify the value
of counters $c_1$ and $c_2$, respectively.
The initial configuration is $(\ell_0, 0, 0)$.
A run of a two-counter machine is a (finite or infinite) sequence of
configurations $\seq{k_0, k_1, \ldots}$ where $k_0$ is the initial
configuration, and the relation between subsequent configurations is
governed by transitions between respective instructions.
The run is a finite sequence if and only if the last configuration is
the terminal instruction $\ell_n$.
Note that a two-counter  machine has exactly one run starting from the initial
configuration. 
The \emph{halting problem} for a two-counter machine asks whether 
its unique run ends at the terminal instruction $\ell_n$.
It is well known~(\cite{Min67}) that the halting problem for
two-counter machines is undecidable.

\subsection{Proofs for undecidability of mean-payoff (per-transition) games}
\label{app:undec}
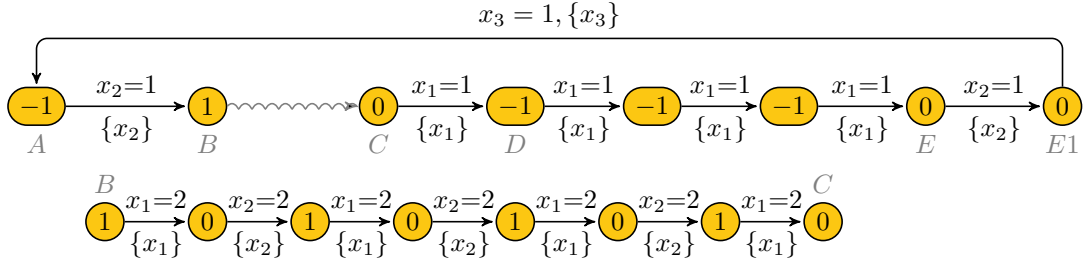
\begin{figure}[t]
\centering
\begin{tikzpicture}[->,>=stealth',shorten >=1pt,auto,node distance=1cm,
  semithick,scale=0.9]

  \node[ player1] at (-3,-3.5)(A) {$-1$};
   \node()[below of=A,node distance=5mm,color=gray]{$A$};

  \node[player1] at (-0.5,-3.5) (B){$1$};
   \node()[below of=B,node distance=5mm,color=gray]{$B$};

  \node[player1] at (2,-3.5) (C){$0$};
   \node()[below of=C,node distance=5mm,color=gray]{$C$};

  \node[player1] at (4,-3.5) (D) {$-1$};
   \node()[below of=D,node distance=5mm,color=gray]{$D$};

   \node[player1] at (6,-3.5) (D1) {$-1$};
   
   \node[player1] at (8,-3.5) (D2) {$-1$};

   \node[player1] at (10,-3.5) (E) {$0$};
   \node()[below of=E,node distance=5mm,color=gray]{$E$};
   
   \node[player1] at (12,-3.5) (E1) {$0$};
   \node()[below of=E1,node distance=5mm,color=gray]{$E1$};

    \path (A) edge node {$ x_2 {=} 1$} node[below]{$\set{x_2}$} (B);
    
    \path (B) edge[gluon,draw=gray]  node {} node[below]{} (C);

    \path (C) edge node {$x_1 {=} 1$} node[below]{$\set{x_1}$} (D);
   
    \path (D) edge node {$x_1 {=} 1$} node[below]{$\set{x_1}$} (D1);
    \path (D1) edge node {$x_1 {=} 1$} node[below]{$\set{x_1}$} (D2);
    \path (D2) edge node {$x_1 {=} 1$} node[below]{$\set{x_1}$} (E);

    \path (E) edge node {$x_2 {=} 1$} node[below]{$\set{x_2}$} (E1);
    \draw[rounded corners] (E1) -- (12, -2.5) -- node[above]{$x_3=1, \{x_3\}$} 
 (-3, -2.5) --  (A);

    \node[player1] at (-2,-5.2) (B1){$1$};
    \node()[above of=B1,node distance=5mm,color=gray]{$B$};
    \node[player1] at (-0.5,-5.2) (C1){$0$};
    \path (B1) edge  node {$x_1 {=} 2$} node[below]{$\set{x_1}$} (C1);
    \node[player1] at (1,-5.2) (C2){$1$};
    \path (C1) edge node[above]{$x_2 {=} 2$} node[below]{$\set{x_2}$} (C2);
    \node[player1] at (2.5,-5.2) (C3){$0$};
    \path (C2) edge node[above]{$x_1 {=} 2$} node[below]{$\set{x_1}$} (C3);
    \node[player1] at (4,-5.2) (C4){$1$};
    \path (C3) edge node[above]{$x_2 {=} 2$} node[below]{$\set{x_2}$} (C4);
    
    \node[player1] at (5.5,-5.2) (C5){$0$};
    \path (C4) edge node[above]{$x_1 {=} 2$} node[below]{$\set{x_1}$} (C5);
    \node[player1] at (7,-5.2) (C6){$1$};
    \path (C5) edge node[above]{$x_2 {=} 2$} node[below]{$\set{x_2}$} (C6);
    \node[player1] at (8.5,-5.2) (C7){$0$};
    \path (C6) edge node[above]{$x_1 {=} 2$} node[below]{$\set{x_1}$} (C7);
    \node()[above of=C7,node distance=5mm,color=gray]{$C$};
    
 \end{tikzpicture}
\caption{$WD^1_1$ redrawn with location prices from $\{1,0,-1\}$. 
Every location has a self loop with the guard $x_2,x_3=1$, reset $x_2,x_3$, which is not shown here 
for conciseness. The curly
 edge from $B$ to  $C$ is shown below.
 The mean-payoff incurred in one transit from $A$  to $A$ via $E$ is $\frac{\epsilon}{14}$. 
If $Min$ makes no error, this is 0.   
}
\label{redraw}
\end{figure}

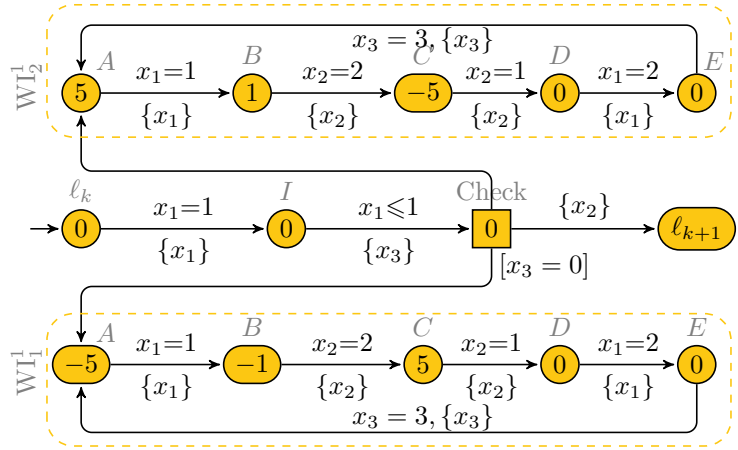
\begin{figure}[tbp]
\centering
\begin{tikzpicture}[->,>=stealth',shorten >=1pt,auto,node distance=1cm,
  semithick,scale=0.9]
  \node[initial,initial text={}, player1] at (0,-.5) (lk) {$0$ } ;
   \node()[above of=lk,node distance=5mm,color=gray]{$\ell_k$};

     \node[player1] at (3,-.5) (i){$0$} ;
     \node()[above of=i,node distance=5mm,color=gray]{$I$};
 
  \node[player2] at (6,-.5) (chk){$0$} ;
  \node()[above of=chk,node distance=5mm,color=gray]{$\text{Check}$};
  \node () [below right of=chk,node distance=7mm,xshift=2mm] {$[x_3=0]$};

   \node[player1] at (9,-.5)(lk1){$\ell_{k+1}$};

\path (lk) edge node {$x_1 {=} 1$} node[below] {$\set{x_1}$}(i);
\path (i) edge node {$x_1 {\leq} 1$} node[below] {$\set{x_3}$}(chk);
\path (chk) edge node[above] {$\set{x_2}$} (lk1);

  \node[ player1] at (0,-2.5)(A) {$-5$};
   \node()[above right of=A,node distance=6mm,xshift=-1mm,color=gray]{$A$};

  \node[player1] at (2.5,-2.5) (B){$-1$};
   \node()[above of=B,node distance=5mm,color=gray]{$B$};

  \node[player1] at (5,-2.5) (C){$5$};
   \node()[above of=C,node distance=5mm,color=gray]{$C$};

  \node[player1] at (7,-2.5) (D) {$0$};
   \node()[above of=D,node distance=5mm,color=gray]{$D$};

  \node[player1] at (9,-2.5) (E) {$0$};
   \node()[above of=E,node distance=5mm,color=gray]{$E$};


  \node () [below of=C,node distance=7mm] {$x_3=3, \{x_3\}$};

  
  \draw[rounded corners] (chk) -- (6, -1.35) -- (0, -1.35) --  (A);

    \path (A) edge node {$ x_1 {=} 1$} node[below]{$\set{x_1}$} (B);
    \path (B) edge node {$x_2 {=} 2$} node[below]{$\set{x_2}$} (C);
    \path (C) edge node {$x_2 {=} 1$} node[below]{$\set{x_2}$} (D);
    \path (D) edge node {$x_1 {=} 2$} node[below]{$\set{x_1}$} (E);
    \draw[rounded corners] (E) -- (9, -3.5) -- (0, -3.5) --  (A);
           
    \node[rotate=90,color=gray] at (-.8, -2.6) (N) {$\text{WI}^1_1$};
    \draw[dashed,draw=gold,rounded corners=10pt] (-.5,-3.7) rectangle (9.5,-1.75);

\node[ player1] at (0,1.5)(A1) {$5$};
   \node()[above right of=A1,node distance=6mm,xshift=-1mm,color=gray]{$A$};

  \node[player1] at (2.5,1.5) (B1){$1$};
   \node()[above of=B1,node distance=5mm,color=gray]{$B$};

  \node[player1] at (5,1.5) (C1){$-5$};
   \node()[above of=C1,node distance=5mm,color=gray]{$C$};

  \node[player1] at (7,1.5) (D1) {$0$};
   \node()[above of=D1,node distance=5mm,color=gray]{$D$};

  \node[player1] at (9,1.5) (E1) {$0$};
   \node()[above right of=E1,node distance=6mm,xshift=-2mm,color=gray]{$E$};


  \node () [above of=C1,node distance=7mm] {$x_3=3, \{x_3\}$};

  \draw[rounded corners] (chk) -- (6, 0.35) -- (0, 0.35) --  (A1);

    \path (A1) edge node {$ x_1 {=} 1$} node[below]{$\set{x_1}$} (B1);
    \path (B1) edge node {$x_2 {=} 2$} node[below]{$\set{x_2}$} (C1);
    \path (C1) edge node {$x_2 {=} 1$} node[below]{$\set{x_2}$} (D1);
    \path (D1) edge node {$x_1 {=} 2$} node[below]{$\set{x_1}$} (E1);

 \draw[rounded corners] (E1) -- (9, 2.5) -- (0, 2.5) --  (A1);

    \node[rotate=90,color=gray] at (-.8, 1.6) (N1) {$\text{WI}^1_2$};
    \draw[dashed,draw=gold,rounded corners=10pt] (-.5,2.8) rectangle (9.5,0.85);

 \end{tikzpicture}
 
\caption{Simulation to increment counter $C_1$, mean cost is $\varepsilon$ for error $\varepsilon$. As seen in Figure \ref{redraw}, the widgets
$\text{WI}^1_2$ and $\text{WI}^1_1$ can be redrawn with location prices in $\{0,1,-1\}$. 
}
\label{fig_undec_mpg_inc_new}
\end{figure}

\paragraph{Simulation of increment instruction}: 
   The module to
  increment $C_1$ is given in Fig.~\ref{fig_undec_mpg_inc_new}. Again,
  we start at $\ell_k$ with $x_1=\frac{1}{5^{c_1}7^{c_2}}, x_2 = 0$ and
  $x_3=0$. Let $\frac{1}{5^{c_1}7^{c_2}}$ be called as $x_{old}$.  A
  time of $1-x_{old}$ is spent at $\ell_k$.  Let the time spent at $I$ be
  denoted $x_{new}$.  To correctly increment counter 1, $x_{new}$ must
  be $\frac{x_{old}}{5}$.  No time is spent at $\text{Check}$. Player
  Max can either continue simulation of the next instruction, or can
  enter one of the widgets $\text{WI}^1_1, \text{WI}^1_2$ to verify if $x_{new}$ is indeed
  $\frac{x_{old}}{5}$.  
  
 If player Min makes an error by elapsing a time $\frac{x_{old}}{5} +\varepsilon$, then 
 player Max chooses the widget $\text{WI}^1_1$. It can be seen that 
 the cost of going from $A$ to $E$ once in $\text{WI}^1_1$ is $5 \varepsilon$. The mean cost 
 incurred by Min in going from $A$ to $A$ through $E$ once, 
 when Max chooses $\text{WI}^1_1$ is hence $\varepsilon$.
 Similarly, if Min makes an error by elapsing a time $\frac{x_{old}}{5} - \varepsilon$, then 
 player Max chooses the widget $\text{WI}^1_2$. 
 The mean cost incurred in one transit from $A$ to $A$ through $E$ in $\text{WI}^1_2$  is again $\varepsilon$.
 If Min makes no simulation error, then the mean cost incurred 
 is 0.

\begin{figure}[tbp]
\begin{center}
\begin{tikzpicture}[->,>=stealth',shorten >=1pt,auto,node distance=1cm,
  semithick,scale=0.8]
  \node[initial,initial text={}, player1] (a) {$0$};
  \node()[above of=a,node distance=5mm,color=gray]{$A$};
  
  \node[player1] at (4,0) (b) {$0$};
  \node()[above of=b,node distance=5mm,color=gray]{$B$};
  
  \node[player2] at (8,0) (chk) {$0$};
  \node()[above of=chk,node distance=5mm,color=gray]{$\text{Check}$};
  \node()[below of=chk,node distance=5mm]{$[x_3{=}0]$};

  \node[widget] at (11,0) (wd11){$\text{WD}^1_1$};
  \node[widget] at (11,2) (wd12){$\text{WD}^1_2$};

  \node[player2] at (0,-3) (d) {$1$};
  \node()[above of=d,node distance=5mm,color=gray]{$D$};
  
  \node[player2] at (0,-5) (t) {$0$};
  \node()[below of=t,node distance=5mm,color=gray]{$T$};
  
%
%
  
  \node[player1] at (4,-3) (t1) {$0$};
  \node()[below of=t1,node distance=5mm,color=gray]{$F$};

  \path (a) edge node {$x_2{=}0$} (b);
  \path (b) edge[bend right=10] node[below] {$x_1< 5, \set{x_3}$} (chk);
  \path (chk) edge [bend right=10] node[above] {$\{x_2\}$} (b);  
  \path (chk) edge node {} (wd11);
  \path (chk) edge node {} (wd12);
  \path (b) edge node[left] {$\begin{array}{c}x_1{=}1 \land\\  x_2=0\end{array}$} (d);
  \path (d) edge[bend left=70] node[left] {$x_2{=}1$} (t);
  \path (d) edge[bend left=70] node[right] {$\set{x_2}$} (t);
  \path (t) edge[bend left=70] node[left] {$x_2=0$} (d);
  
  \path (b) edge node[right]{$\begin{array}{c}x_1{>}1\land \\ x_2=0\end{array}$} (t1);

  \path (t1) edge [loop right] node {$x_2=0$} (t1);
\end{tikzpicture}
\caption{ Widget $\text{W}_2^{\neq 0}$}
\label{fig_undec_mpg_wzne_new}
\end{center}
\end{figure}
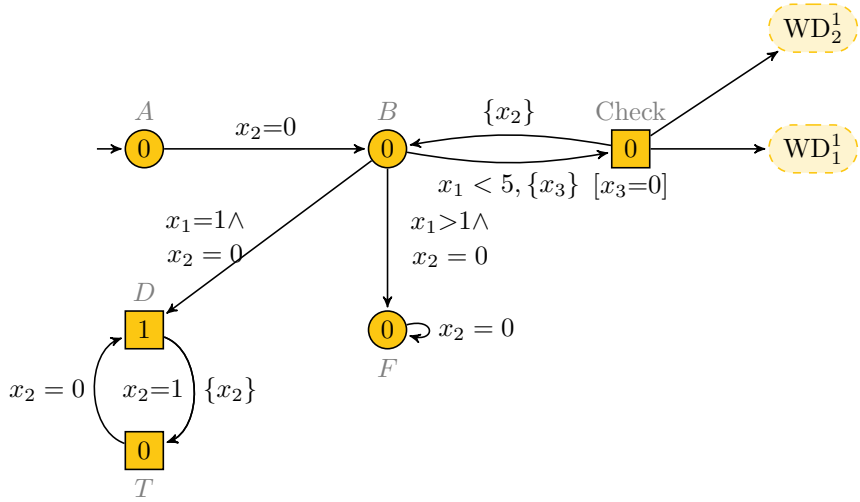

  \paragraph{Simulation of Zero-check}:
  Fig.~\ref{fig_undec_mpg_zeroCheck} shows the module for zero-check
  instruction for counter $C_2$. $\ell_k$ is a no time elapse location,
  from where, player Min chooses one of the locations $\text{Check}_{c_2=0}$ or
  $\text{Check}_{c_2 \neq 0}$. Both these are player Max locations, and player Max
  can either continue the simulation, or can go to the check widgets
  $\text{W}_2^{=0}$ or $\text{W}_2^{\neq 0}$ to verify the correctness of player Min's choice.
  The widgets $\text{W}_2^{=0}$ and $\text{W}_2^{\neq 0}$ are given in
  Fig.~\ref{fig_undec_mpg_wze} and Fig. \ref{fig_undec_mpg_wzne_new}
  respectively.
\begin{figure}[tbp]
\centering
    \begin{tikzpicture}[->,>=stealth',shorten >=1pt,auto,node distance=1cm,
      semithick,scale=0.8]
      \node[initial,initial text={}, player1] at (-.5,0) (A) {$0$};
      \node()[above of=A,node distance=5mm,color=gray]{$\ell_k$};
      \node () [below right of=A,node distance=6.5mm,xshift=-3mm] {$[x_3=0]$};

      \node[player2] at (2,1) (B){$0$} ;
       \node()[below of=B,node distance=5mm,color=gray]{$\text{Check}_{c_2=0}$};
       
      \node[widget] at (2,2.5) (B1){$\text{W}_2^{= 0}$ };

      \node[player2] at (2,-1) (C){$0$} ;
      \node()[above of=C,node distance=5mm,color=gray]{$\text{Check}_{c_2\neq 0}$};
      
      \node[widget] at (2,-2.5) (C1){$\text{W}_2^{\neq 0}$};

      \node[player1] at (4.5,1) (D) {$0$};
      \node()[above of=D,node distance=5mm,color=gray]{$\ell_{k+1}^1$};

      \node[player1] at (4.5,-1) (E) {$0$ };
      \node()[above of=E,node distance=5mm,color=gray]{$\ell_{k+1}^2$};

      \path (A) edge (B);
      \path (A) edge (C);
      \path (B) edge node {$x_3=0$} (B1);
      \path (B) edge node[above] {$x_3=0$} (D);
      \path (C) edge node[left] {$x_3=0$} (C1);
      \path (C) edge node[above] {$x_3=0$} (E);
    \end{tikzpicture}
    \caption{Widget $\text{WZ}_2$ simulating
      zero-check for $C_2$}

\label{fig_undec_mpg_zeroCheck}
\end{figure}
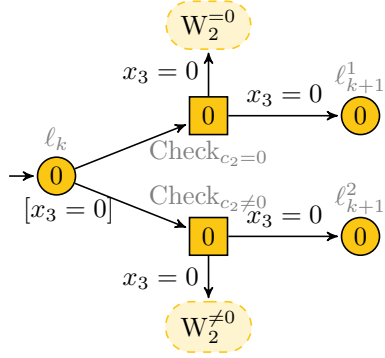
\begin{figure}[tbp]
 \begin{center}
\begin{tikzpicture}[->,>=stealth',shorten >=1pt,auto,node distance=1cm,
  semithick,scale=0.8]
  \node[initial,initial text={}, player1] at (1,0) (a) {$0$};
  \node()[above of=a,node distance=5mm,color=gray]{$A$};
  
  \node[player1] at (4,0) (b) {$0$};
  \node()[above of=b,node distance=5mm,color=gray]{$B$};
  
  \node[player2] at (8,0) (chk) {$0$};
  \node()[above of=chk,node distance=5mm,color=gray]{Check};
  \node()[below right of=chk,node distance=6.5mm,xshift=1mm]{$[x_3{=}0]$};

  \node[widget] at (11,2) (wd11){$\text{WD}^1_1$};
  \node[widget] at (11,0) (wd12){$\text{WD}^1_2$};

  \node[player2] at (8,-3) (c) {$1$};
  \node()[below of=c,node distance=5mm,color=gray]{C};
  
  \node[player2] at (11,-3) (d) {$0$};
  \node()[below of=d,node distance=5mm,color=gray]{$T$};
  
  \node[player1] at (4,-3) (t) {$0$};
  \node()[below of=t,node distance=5mm,color=gray]{$F$};
  
  \path (a) edge node {$x_2{=}0$} (b);
  \path (b) edge[bend right=10] node[below] {$\set{x_3}$} (chk);
  \path (chk) edge [bend right=10] node[above] {$x_1 {\leq} 1,\{x_2\}$} (b);  
  \path (chk) edge node {} (wd11);
  \path (chk) edge node {} (wd12);
  \path (b) edge node[left] {$\begin{array}{c}x_1{=}1\\ \land x_2=0\end{array}$} (t);
  \path (t) edge [loop left] node {$x_2=0$} (t);
  \path (chk) edge node[left]{$x_1{>}1$} node[right]{$\set{x_2}$} (c);
  \path (c) edge[bend right=20] node[above] {$x_2=1$} (d);
  \path (c) edge[bend right=20] node[below] {$\set{x_2}$} (d);
  \path (d) edge[bend right=20] node[above] {$x_2=0$} (c);

\end{tikzpicture}

  \end{center}
 \caption{ Widget $\text{W}_2^{=0}$}
\label{fig_undec_mpg_wze}
\end{figure}
               
  \begin{itemize}
  \item Consider the case when player Min guessed that $C_2$ is zero,
    and entered the location $\text{Check}_{c_2=0}$ in
    Fig.~\ref{fig_undec_mpg_zeroCheck}.  Let us assume that player Max
    verifies player Min's guess by entering $\text{W}_2^{=0}$
    (Fig.~\ref{fig_undec_mpg_wze}).  No time is spent in the initial
    location $A$ of $\text{W}_2^{=0}$.  We are therefore at $B$ with
    $x_1=\frac{1}{5^{c_1}7^{c_2}}=x_{old}$ and $x_2,x_3=0$.  In case
    $c_1=c_2=0$, we can directly go to the $F$ state, and keep looping there
    forever, incurring mean cost 0.  If that is not the case, player Min has
    to prove his claim right, by multiplying $x_1$ with 5 repeatedly,
    till $x_1$ becomes 1; clearly, this is possible iff $c_2=0$.  The
    loop between $B$ and $\text{Check}$ precisely does this: each time
    player Min spends a time $x_{delay}$ in $B$, player Max can verify that
    $x_{delay}=4 x_{old}$ by going to $\text{WD}^1_1$ or $\text{WD}^1_2$ or come back to
    $B$.  No time is elapsed in $\text{Check}$. Finally, if $x_1=1$,
    we can go to $F$, and player Min achieves his objective. However, if
    $C_2$ was non-zero, then $x_1$ will never reach 1 after repeatedly
    multiplying $x_1$ with 5; in this case, at some point, the edge
    from $\text{Check}$ to $C$ will be enabled. In this case, the
    infinite loop between $C$ and $T$, 
   will lead to a mean cost greater than 0.

  \item
  Consider now the case when player Min guessed that $C_2$ is
    non-zero, and hence entered the location $\text{Check}_{c_2\neq
      0}$ in Fig.~\ref{fig_undec_mpg_zeroCheck}.  Let us assume now
    that player Max enters $\text{W}_2^{\neq 0}$
    (Fig.~\ref{fig_undec_mpg_wzne_new}) to verify player Min's guess. 
    Similar to $\text{W}_2^{=0}$, no time is spent at location
    $A$ of $\text{W}_2^{\neq 0}$, and the clock values at $B$ are
    $x_1=\frac{1}{5^{c_1}7^{c_2}}=x_{old}$ and $x_2,x_3=0$.  
    If $c_1=c_2=0$, then $x_1=1$, in which case, the location $D$ is
    reached, from where, the loop between $D,T$ is taken 
    incurring a mean
    cost greater than 0.

	If that is not the case, player $\mMIN$, to prove his claim,
	repeatedly multiplies $x_1$ by 5 using the
	loop between $B$ and $\text{Check}$.
	$x_1$ becomes $1$ iff $c_2=0$.
	Once $x_1$ becomes $1$, the edge
    from $B$ to $D$ will be enabled. In this case, the
    infinite loop between $D$ and $T$, 
   will lead to a mean cost greater than 0.
   Note that once $x_1$ becomes 1, player $\mMIN$ can also wait in $B$
   and transit to $\text{Check}$.
   However, due to the guard on the edge from $B$ to $\text{Check}$,
   the delay at $B$ will be less than $4x_{old}$, (when $x_1=1$, $x_1=x_{old}=\frac{1}{5^{c_1}7^{c_2}}=1$,  
$c_1=c_2=0$)    say $4x_{old} - \varepsilon$
   in which case too player $\mMAX$ can go to $\text{WD}^1_1$ or $\text{WD}^1_2$
   and the mean cost will be $\varepsilon$.
    If $C_2$ was non-zero, then $x_1$ will never reach 1 after repeatedly
    multiplying $x_1$ with 5; in this case, at some point, 
	$x_1$ will be greater than 1 and
    the edge
    from $B$ to $F$ will be enabled and player $\mMIN$
    can achieve his objective by moving to $F$.
      \end{itemize}
  
  \paragraph{Correctness of the construction}
  On entry into the location $\location_n$ (for HALT instruction), we
  reset clock $x_1$ to 0; from $\location_n$, we go to a state $F$
  with price 1, with a self loop that checks $x_1=1$, and resets $x_1$.
  \begin{enumerate}
  \item Assume that the two counter machine halts. If player Min
    simulates all the instructions correctly, he will incur a mean cost
    $>0$, by either reaching the $F$ after $\location_n$ :
     when player Min does not cheat, player Max has no incentive to enter any of the check widgets,
     he will just let the computation continue, till the Halt  location is reached. This will incur a mean cost 
     $>0$. 
     If player Min makes an error in his computation, player
    Max can always enter an appropriate widget, making the mean cost $>0$.   In summary, if the two counter machine halts, then  player Min has
    no strategy to achieve his goal (mean pay off $\leq 0$).
  \item Assume that the two counter machine does not halt.
    \begin{itemize}
    \item If player Min simulates all the instructions correctly, and if
      player Max never enters a check widget, then player Min incurs cost
      0, since all locations in the main modules have price 0. 
      Even if player Max enters some widget, the mean cost of player Min 
      is still 0, since no errors were made by Min. 
      However, if player Min makes an error, player Max can enter a check widget, ensuring that 
       the mean cost is $>0$. Thus, if the two counter machine does not halt, player Min has a strategy 
       (by making no errors) to achieve mean cost 0.
            \end{itemize}
    In summary, if the two counter machine does not halt, player Min
    has a strategy to achieve his goal, and vice-versa.
  \end{enumerate}
  Thus, player Min incurs a mean cost $\leq 0$ iff he
  chooses the strategy of faithfully simulating the two counter
  machine, when the machine does not halt. When the machine  halts, the
  mean cost incurred by player Min is more than 0
  irrespective 
  of whether he makes a simulation error or not.  \qed

\subsection*{$\MPG(\pta, r)$ problem, $r>0$} 
  Now we argue that the $\MPG(\pta, r)$ problem is undecidable for 
  PTGAs with $\geq 3$  clocks, having only 
  binary location prices and no edge prices, for $r=\frac{1}{3}$. 
  We note that in the previous undecidability result,
  all the modules apart from the $WD^1_1, WD^1_2, WI^1_1$ and $WI^1_2$
  use binary location prices.
  Thus we now give here these modules with only binary prices.
  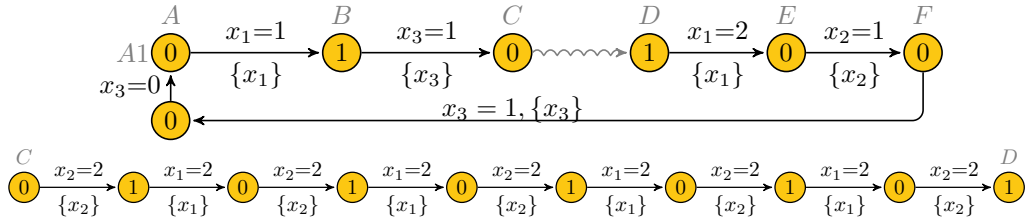
\begin{figure}[h]
\centering
\begin{tikzpicture}[->,>=stealth',shorten >=1pt,auto,node distance=1cm,
  semithick,scale=0.9]
\tikzset{every loop/.style={min distance=15mm}}

  \node[ player1] at (0,-3.5)(A) {$0$};
   \node()[above of=A,node distance=5mm,color=gray]{$A$};

  \node[ player1] at (0,-4.5)(A1) {$0$};
   \node()[left of=A,node distance=5mm,color=gray]{$A1$};

  \node[player1] at (2.5,-3.5) (B){$1$};
   \node()[above of=B,node distance=5mm,color=gray]{$B$};

  \node[player1] at (5,-3.5) (C){$0$};
   \node()[above of=C,node distance=5mm,color=gray]{$C$};

  \node[player1] at (7,-3.5) (D) {$1$};
   \node()[above of=D,node distance=5mm,color=gray]{$D$};

  \node[player1] at (9,-3.5) (E) {$0$};
   \node()[above of=E,node distance=5mm,color=gray]{$E$};

   \node[player1] at (11,-3.5) (F) {$0$};
   \node()[above of=F,node distance=5mm,color=gray]{$F$};

   \node () [below of=C,node distance=7.5mm] {$x_3=1, \{x_3\}$};

    \path (A) edge node {$ x_1 {=} 1$} node[below]{$\set{x_1}$} (B);
    \path (B) edge node {$x_3 {=} 1$} node[below]{$\set{x_3}$} (C);
    \path (C) edge[gluon,gray] node{} (D);
    \path (D) edge node {$x_1 {=} 2$} node[below]{$\set{x_1}$} (E);
    
    \path (E) edge node {$x_2 {=} 1$} node[below]{$\set{x_2}$} (F);
    \draw[rounded corners] (F) -- (11, -4.5)  --  (A1);
    \path (A1) edge node {$ x_3 {=} 0$} (A);

\end{tikzpicture}
\scalebox{0.8}{
\begin{tikzpicture}[->,>=stealth',shorten >=1pt,auto,node distance=1cm,
  semithick,scale=0.9]
\tikzset{every loop/.style={min distance=15mm}}
    \node[player1] at (-6,-5.5) (C1){$0$};
    \node()[above of=C1,node distance=5mm,color=gray]{$C$};
    \node[player1] at (-4,-5.5) (D1) {$1$};
    \path (C1) edge node {$x_2 {=} 2$} node[below]{$\set{x_2}$} (D1);
    \node[player1] at (-2,-5.5) (C2){$0$};
    \path (D1) edge node[above]{$x_1 {=} 2$} node[below]{$\set{x_1}$} (C2);
    \node[player1] at (0,-5.5) (D2) {$1$};
    \path (C2) edge node {$x_2 {=} 2$} node[below]{$\set{x_2}$} (D2);
    \node[player1] at (2,-5.5) (C3){$0$};
    \path (D2) edge node[above]{$x_1 {=} 2$} node[below]{$\set{x_1}$} (C3);
    \node[player1] at (4,-5.5) (D3) {$1$};
    \path (C3) edge node {$x_2 {=} 2$} node[below]{$\set{x_2}$} (D3);
    \node[player1] at (6,-5.5) (C4){$0$};
    \path (D3) edge node[above]{$x_1 {=} 2$} node[below]{$\set{x_1}$} (C4);
    \node[player1] at (8,-5.5) (D4) {$1$};
    \path (C4) edge node {$x_2 {=} 2$} node[below]{$\set{x_2}$} (D4);
    
    \node[player1] at (10,-5.5) (C5){$0$};
    \path (D4) edge node[above]{$x_1 {=} 2$} node[below]{$\set{x_1}$} (C5);
    \node[player1] at (12,-5.5) (D5) {$1$};
    \path (C5) edge node {$x_2 {=} 2$} node[below]{$\set{x_2}$} (D5);
    \node()[above of=D5,node distance=5mm,color=gray]{$D$};
\end{tikzpicture}
}
\caption{$WD^1_1$ here has locations 
  with binary price rates. All locations have a self loop $x_3=1?$, reset $x_3$. 
  This module is chosen by the $\mMAX$ player
  when the error made by $\mMIN$ player
  during the decrement operation is $\varepsilon$.
   The curly edge from $C$ to  $D$ is shown by the path below. The mean-payoff
   incurred in one transit from $A$  to $A$ via $F$ is $\frac{1}{3} +
   \frac{\epsilon}{15}$. 
  If $\mMIN$ makes no error, the cost is $\frac{1}{3}$.
}
\label{dec_binary_pos}
\end{figure}
\begin{figure}[h]
  \centering
  \begin{tikzpicture}[->,>=stealth',shorten >=1pt,auto,node distance=1cm,
      semithick,scale=0.9]
    \tikzset{every loop/.style={min distance=15mm}}
    
    \node[ player1] at (0,-0)(A) {$1$};
    \node()[above of=A,node distance=5mm,color=gray]{$A$};
    
    \node[ player1] at (4,-1)(A1) {$0$};
    \node()[node distance=5mm,color=gray]{};

    \node[ player1] at (2,-1)(A2) {$0$};
    \node()[node distance=5mm,color=gray]{};
    
    \node[ player1] at (0,-1)(A3) {$0$};
    \node()[node distance=5mm,color=gray]{};
    
    \node[player1] at (2.5,0) (B){$0$};
    \node()[above of=B,node distance=5mm,color=gray]{$B$};
    
    \node[player1] at (5,0) (C){$1$};
    \node()[above of=C,node distance=5mm,color=gray]{$C$};

    \node[player1] at (7,0) (D) {$0$};
    \node()[above of=D,node distance=5mm,color=gray]{$D$};
    
    \node[player1] at (9,0) (E) {$0$};
    \node()[above of=E,node distance=5mm,color=gray]{$E$};
    
    \node[player1] at (11,0) (F) {$0$};
    \node()[above of=F,node distance=5mm,color=gray]{$F$};
    \node () [below of=D,node distance=7.5mm] {$x_3=1, \{x_3\}$};

    \path (A) edge node {$ x_2 {=} 1$} node[below]{$\set{x_2}$} (B);
    \path (B) edge[gluon,gray] node{} (C);
    \path (C) edge node {$x_2 {=} 2$} node[below]{$\set{x_2}$} (D);
    \path (D) edge node {$x_1 {=} 2$} node[below]{$\set{x_1}$} (E);
    
    \path (E) edge node {$x_2 {=} 1$} node[below]{$\set{x_2}$} (F);
    \draw[rounded corners] (F) -- (11, -1)--  (A1);
    \path (A1) edge node {$ x_3 {=} 0$}  (A2);
    \path (A2) edge node {$ x_3 {=} 0$}  (A3);
    \path (A3) edge node {$ x_3 {=} 0$}  (A);
\end{tikzpicture}
\scalebox{0.9}{
    
\begin{tikzpicture}[->,>=stealth',shorten >=1pt,auto,node distance=1cm,
  semithick,scale=0.9]

 \node[player1] at (-2,-6.2) (B1){$0$};
   \node()[above of=B1,node distance=5mm,color=gray]{$B$};
 \node[player1] at (-0.5,-6.2) (C1){$1$};
\path (B1) edge  node {$x_1 {=} 2$} node[below]{$\set{x_1}$} (C1);
 \node[player1] at (1,-6.2) (C2){$0$};
\path (C1) edge node[above]{$x_2 {=} 2$} node[below]{$\set{x_2}$} (C2);
 \node[player1] at (2.5,-6.2) (C3){$1$};
\path (C2) edge node[above]{$x_1 {=} 2$} node[below]{$\set{x_1}$} (C3);
\node[player1] at (4,-6.2) (C4){$0$};
\path (C3) edge node[above]{$x_2 {=} 2$} node[below]{$\set{x_2}$} (C4);

 \node[player1] at (5.5,-6.2) (C5){$1$};
\path (C4) edge node[above]{$x_1 {=} 2$} node[below]{$\set{x_1}$} (C5);
\node[player1] at (7,-6.2) (C6){$0$};
\path (C5) edge node[above]{$x_2 {=} 2$} node[below]{$\set{x_2}$} (C6);
\node[player1] at (8.5,-6.2) (C7){$1$};
\path (C6) edge node[above]{$x_1 {=} 2$} node[below]{$\set{x_1}$} (C7);
\node()[above of=C7,node distance=5mm,color=gray]{$C$};

 \end{tikzpicture}
}\caption{$WD^1_2$ here has locations 
with binary price rates.
 All locations have a self loop $x_3=1?$, reset $x_3$. 
This module is chosen by the $\mMAX$ player
when the error made by $\mMIN$ player
during the decrement operation is $-\varepsilon$.
The curly edge from $C$ to  $D$ is shown by the path below. The mean-payoff incurred in one transit from $A$  to $A$ via $F$ is $\frac{1}{3} + \frac{\epsilon}{15}$.
If $\mMIN$ makes no error, it is $\frac{1}{3}$.
}
\label{dec_binary_neg}
\end{figure}
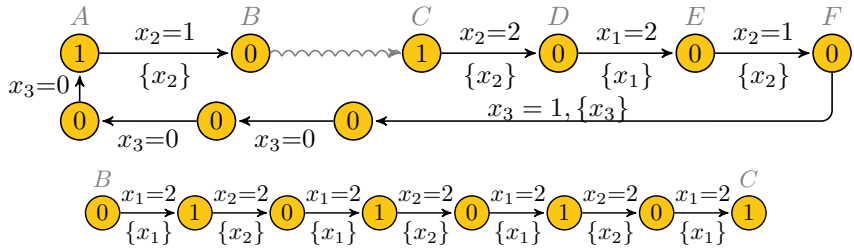

  Fig. \ref{dec_binary_pos} shows the module $WD^1_1$
  while Fig. \ref{dec_binary_neg} shows the module $WD^1_2$
  with location prices 0 and 1.
  The modules $WI^1_1$ and $WI^1_2$ can also be similarly
  made with only binary prices.
  The module $WD^1_1$ is chosen by the $\mMAX$ player
  when the error made by the $\mMIN$ player while
  simulating the decrement counter $C_1$ operation is $\varepsilon$.
  The module $WD^1_2$ is chosen by the $\mMAX$ player
  when the error made by the $\mMIN$ player is $-\varepsilon$.  
  When no error is made by the $\mMIN$ player, 
  the mean cost $WD^1_1$ is $\frac{1}{3}$
  while the cost in the presence of an error is $\frac{1}{3}+ \frac{\epsilon}{15}$.

The modules for zerocheck (Figure \ref{fig_undec_mpg_zeroCheck}) can be used as they are. In case an error is made in the guess, the meanpayoff 
incurred by shuttling between locations $C,T$ (Figure \ref{fig_undec_mpg_wze}
 ) or $D,T$ (Figure \ref{fig_undec_mpg_wzne_new}
) is $\frac{1}{2} > \frac{1}{3}$. 

\subsection{Proofs for undecidability of mean-payoff (per-time-unit) games}

\label{app:undec_per_time_unit}
Mean-payoff game (per time unit) has been studied in \cite{BCR14}.
The mean-payoff (per time unit) of a play is defined as the long-run average
of cost per time unit.
Formally, the mean payoff of a play 
$\RUN(s,\mu.\chi) = \sigma = (\ell_0, \nu_0)\trans{(d_0, a_0)} (\ell_1,
\nu_1)\trans{(d_1, a_1)} \cdots$ starting from a state $s$  
in which players $\mMin$ and $\mMax$ play according to
$\mu$ and $\chi$ respectively is
\[
\lim_{n \rightarrow \infty}
\frac{
  \sum_{i=0}^n  (p(\ell_i,a) + p(\ell_1) \cdot t_i)}{
  \sum_{i=0}^n t_i}.
\]

We refine the  undecidability result in \cite{BCR14} by
showing the following result. 
\begin{theorem} \label{thm-undec_per_time}
The mean-payoff problem $\MPG(\pta, r)$ is undecidable for PTGA
$\pta$ with 3 clocks having location-wise  price-rates $\pi(\ell) \in \{0,1,-1\}$ for
all $\ell \in L$ and $r = 0$.
Moreover, it is undecidable for binary-priced $\pta$ with 3 clocks and $r >0$.
\end{theorem}
Similar to the proof of Theorem \ref{thm_undec},
this proof also involves reduction from the non-halting problem of two counter machines.
\begin{proof}
We first show the undecidability of $\MPG(\pta, 0)$ with location prices $\{1,0,-1\}$ and no edge prices.
We prove the result by reducing the non-halting problem 
of  2 counter machines. Given a two counter machine ${\cal M}$, we 
construct a PTGA $\Gamma$ with 3 clocks $x_1, x_2, x_3$, and arbitrary location prices, but no edge prices, 
and show that player $\mMIN$ has a  winning strategy 
iff ${\cal M}$ does not halt. 
      
  We specify a module for each instruction of the two counter machine.
  On entry into a module, we have $x_1 =
  \frac{1}{5^{c_1}7^{c_2}}$, $x_2 = 0$ and $x_3=0$, where $c_1,c_2$ are
  the values of counters $C_1,C_2$. We construct the PTGA $\Gamma$ whose building
  blocks are the modules for instructions. 
  The role of player Min is to
  faithfully simulate the two counter machine, by choosing appropriate
  delays to adjust the clocks to reflect changes in counter values.
  Player Max will have the opportunity to verify that player $\mMIN$ did not
  cheat while simulating the machine.  We shall now present modules
  for increment, decrement and zero check instructions. For conciseness of the 
  figures, we present here modules using arbitrary prices. However, 
  we can redraw these with extra locations and edges using only the location prices 
  from $\{1,0,-1\}$, as was done in the case of Theorem \ref{thm_undec}.

 
  \begin{figure}[tbp]
\centering
\begin{tikzpicture}[->,>=stealth',shorten >=1pt,auto,node distance=1cm,
  semithick,scale=0.9]
  \node[initial,initial text={}, player1] at (0,-.5) (lk) {$0$ } ;
   \node()[above of=lk,node distance=5mm,color=gray]{$\ell_k$};

  \node[player2] at (3,-.5) (chk){$0$} ;
     \node()[above of=chk,node distance=5mm,color=gray]{$\text{Check}$};

  \node () [below right of=chk,node distance=6.5mm,xshift=-3mm] {$[x_3=0]$};

  \node[player1] at (6,-.5) (lk1) {$0$ } ;
   \node()[above of=lk1,node distance=5mm,color=gray]{$\ell_{k+1}$};


\node[widget] at (0,-2) (W1){$WD^1_1$};

\node[widget] at (6,-2) (W2){$WD^1_2$};

  \node[ player1] at (0,-3.5)(A) {$-5$};
   \node()[above of=A,node distance=5mm,color=gray]{$A$};

  \node[player1] at (2.5,-3.5) (B){$20$};
   \node()[above of=B,node distance=5mm,color=gray]{$B$};

  \node[player1] at (5,-3.5) (C){$-15$};
   \node()[above of=C,node distance=5mm,color=gray]{$C$};

  \node[player1] at (7,-3.5) (D) {$0$};
   \node()[above of=D,node distance=5mm,color=gray]{$D$};

  \node[player1] at (9,-3.5) (E) {$0$};
   \node()[above of=E,node distance=5mm,color=gray]{$E$};


  \node () [below of=C,node distance=7mm] {$x_3=3, \{x_3\}$};
    
\path (lk) edge node {$x_1 {\leq} 1$} node[below] {$\set{x_3}$}(chk);
\path (chk) edge node[below] {$\set{x_2}$} (lk1);
\path (chk) edge node {} (W1);
\path (chk) edge node {} (W2);

    \path (A) edge node {$ x_2 {=} 1$} node[below]{$\set{x_2}$} (B);
    \path (B) edge node {$x_1 {=} 2$} node[below]{$\set{x_1}$} (C);
    \path (C) edge node {$x_1 {=} 1$} node[below]{$\set{x_1}$} (D);
    \path (D) edge node {$x_2 {=} 2$} node[below]{$\set{x_2}$} (E);
    \draw[rounded corners] (E) -- (9, -4.5) -- (0, -4.5) --  (A);
           
    \node[rotate=90,color=gray] at (-.8, -3.5) (N) {$\text{WD}^1_1$};
    \draw[dashed,draw=gold,rounded corners=10pt] (-.5,-4.7) rectangle (9.5,-2.75);

  \node[ player1] at (0,-6)(A1) {$5$};
   \node()[above of=A1,node distance=5mm,color=gray]{$A$};

  \node[player1] at (2.5,-6) (B1){$-20$};
   \node()[above of=B1,node distance=5mm,color=gray]{$B$};

  \node[player1] at (5,-6) (C1){$15$};
   \node()[above of=C1,node distance=5mm,color=gray]{$C$};

  \node[player1] at (7,-6) (D1) {$0$};
   \node()[above of=D1,node distance=5mm,color=gray]{$D$};

  \node[player1] at (9,-6) (E1) {$0$};
   \node()[above of=E1,node distance=5mm,color=gray]{$E$};


  \node () [below of=C1,node distance=7mm] {$x_3=3, \{x_3\}$};
    ;

    \path (A1) edge node {$ x_2 {=} 1$} node[below]{$\set{x_2}$} (B1);
    \path (B1) edge node {$x_1 {=} 2$} node[below]{$\set{x_1}$} (C1);
    \path (C1) edge node {$x_1 {=} 1$} node[below]{$\set{x_1}$} (D1);
    \path (D1) edge node {$x_2 {=} 2$} node[below]{$\set{x_2}$} (E1);
    \draw[rounded corners] (E1) -- (9, -7) -- (0, -7) --  (A1);
           
    \node[rotate=90,color=gray] at (-.8, -6) (N) {$\text{WD}^1_2$};
    \draw[dashed,draw=gold,rounded corners=10pt] (-.5,-7.2) rectangle (9.5,-5.25);

 \end{tikzpicture}
\caption{Simulation to decrement counter $C_1$, mean cost is $\frac{5\varepsilon}{3}$ for error $\varepsilon$}
\label{fig-undec_mpg_dec_new_ptu}
\end{figure}
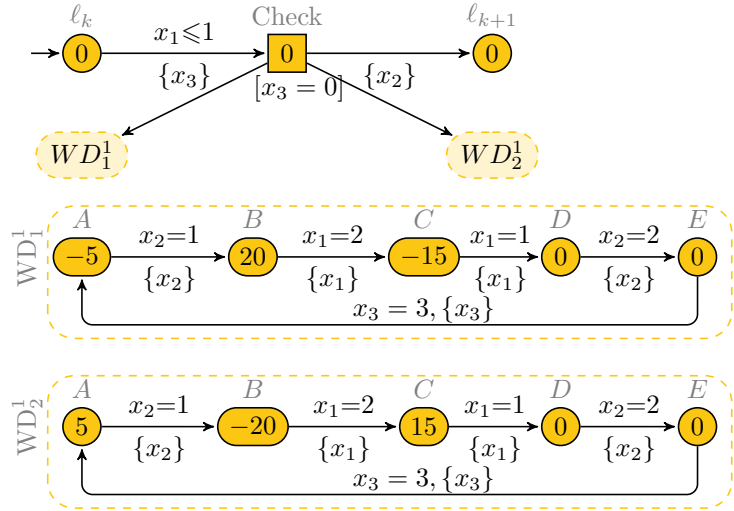

  \paragraph{Simulation of decrement instruction}
  The module to
  simulate the decrement of counter $C_1$ is given in
  Figure~\ref{fig-undec_mpg_dec_new_ptu}.  We enter location $\ell_k$ with
  $x_1=\frac{1}{5^{c_1}7^{c_2}}$, $x_2 = 0$ and $x_3=0$. Lets denote by
  $x_{old}$ the value $\frac{1}{5^{c_1}7^{c_2}}$.  To correctly
  decrement $C_1$, player Min should choose a delay of $4x_{old}$ at
  location $\ell_k$.  At location $\text{Check}$, there is no time
  elapse. Player Max has three possibilities : ($i$) to go to
  $\ell_{k+1}$, or ($ii$) to enter the widget $\text{WD}^1_1$, or (iii) to enter the widget $\text{WD}^1_2$.  If
  player Min makes an error, and delays $4x_{old}+\varepsilon$ 
  or $4x_{old}-\varepsilon$ at
  $\ell_k$ ($\varepsilon > 0$), then player Max can enter one of the widgets
   and punish player Min. Player Max enters widget $\text{WD}^1_1$ if 
     the error made by player Min is of the form 
   $4x_{old}+\varepsilon$ at $\ell_k$ 
  and   enters widget $\text{WD}^1_2$ if the error made by player Min 
  is of the form $4x_{old}-\varepsilon$ at
  $\ell_k$. 
  Let us examine the widget $\text{WD}^1_1$.
   
   When we enter $\text{WD}^1_1$ for
  the first time, we have $x_1=x_{old}+4x_{old}+\varepsilon$,
  $x_2=4x_{old}+\varepsilon$ and $x_3=0$.  In $\text{WD}^1_1$, the cost
  of going once from location $A$ to $E$ is $5\varepsilon$. Also, when we
  get back to $A$ after going through the loop once, the clock values
  with which we entered $\text{WD}^1_1$ are restored; thus, each time,
  we come back to $A$, we restore the starting values with which we
  enter $\text{WD}^1_1$.  The third clock is really useful for this
  purpose only.
  The total time elapsed in 3 time units;
  at $A$, the value of clock $x_3$ is 0 and the transition from 
  $E$ to $A$ is enabled when $x_3=3$.
  Hence the mean cost of 
  transiting from $A$ to $A$ through $E$ is $\frac{5\varepsilon}{3}$. 
  In a similar way, it can be checked 
  that the mean cost of transiting from $A$ to $A$ through $E$
  in widget $\text{WD}^1_2$ is $\frac{5\varepsilon}{3}$ when player Min chooses a delay 
  $4x_{old}-\varepsilon$ at $\ell_k$. 
    Thus, if player Min makes a simulation error, player 
Max can always choose to go to one of the widgets, and ensure that the mean pay-off is \emph{not} $\leq 0$.  
        Note however that when $\varepsilon=0$, then player Min will always achieve his
  objective: the mean pay-off will be 0.

  \begin{figure}[h]
\centering
\begin{tikzpicture}[->,>=stealth',shorten >=1pt,auto,node distance=1cm,
  semithick,scale=0.9]
  \node[initial,initial text={}, player1] at (0,-.5) (lk) {$0$ } ;
   \node()[above of=lk,node distance=5mm,color=gray]{$\ell_k$};

     \node[player1] at (3,-.5) (i){$0$} ;
     \node()[above of=i,node distance=5mm,color=gray]{$I$};
 
  \node[player2] at (6,-.5) (chk){$0$} ;
  \node()[above of=chk,node distance=5mm,color=gray]{$\text{Check}$};
  \node () [below right of=chk,node distance=7mm,xshift=2mm] {$[x_3=0]$};

  \node[player1] at (9,-.5) (lk1) {$0$ } ;
   \node()[above of=lk1,node distance=5mm,color=gray]{$\ell_{k+1}$};

\path (lk) edge node {$x_1 {=} 1$} node[below] {$\set{x_1}$}(i);
\path (i) edge node {$x_1 {\leq} 1$} node[below] {$\set{x_3}$}(chk);
\path (chk) edge node[above] {$\set{x_2}$} (lk1);

  \node[ player1] at (0,-2.5)(A) {$-5$};
   \node()[above right of=A,node distance=6mm,xshift=-1mm,color=gray]{$A$};

  \node[player1] at (2.5,-2.5) (B){$-1$};
   \node()[above of=B,node distance=5mm,color=gray]{$B$};

  \node[player1] at (5,-2.5) (C){$5$};
   \node()[above of=C,node distance=5mm,color=gray]{$C$};

  \node[player1] at (7,-2.5) (D) {$0$};
   \node()[above of=D,node distance=5mm,color=gray]{$D$};

  \node[player1] at (9,-2.5) (E) {$0$};
   \node()[above of=E,node distance=5mm,color=gray]{$E$};


  \node () [below of=C,node distance=7mm] {$x_3=3, \{x_3\}$};

  
  \draw[rounded corners] (chk) -- (6, -1.35) -- (0, -1.35) --  (A);

    \path (A) edge node {$ x_1 {=} 1$} node[below]{$\set{x_1}$} (B);
    \path (B) edge node {$x_2 {=} 2$} node[below]{$\set{x_2}$} (C);
    \path (C) edge node {$x_2 {=} 1$} node[below]{$\set{x_2}$} (D);
    \path (D) edge node {$x_1 {=} 2$} node[below]{$\set{x_1}$} (E);
    \draw[rounded corners] (E) -- (9, -3.5) -- (0, -3.5) --  (A);
           
    \node[rotate=90,color=gray] at (-.8, -2.6) (N) {$\text{WI}^1_1$};
    \draw[dashed,draw=gold,rounded corners=10pt] (-.5,-3.7) rectangle (9.5,-1.75);

\node[ player1] at (0,1.5)(A1) {$5$};
   \node()[above right of=A1,node distance=6mm,xshift=-1mm,color=gray]{$A$};

  \node[player1] at (2.5,1.5) (B1){$1$};
   \node()[above of=B1,node distance=5mm,color=gray]{$B$};

  \node[player1] at (5,1.5) (C1){$-5$};
   \node()[above of=C1,node distance=5mm,color=gray]{$C$};

  \node[player1] at (7,1.5) (D1) {$0$};
   \node()[above of=D1,node distance=5mm,color=gray]{$D$};

  \node[player1] at (9,1.5) (E1) {$0$};
   \node()[above right of=E1,node distance=6mm,xshift=-2mm,color=gray]{$E$};


  \node () [above of=C1,node distance=7mm] {$x_3=3, \{x_3\}$};

  \draw[rounded corners] (chk) -- (6, 0.35) -- (0, 0.35) --  (A1);

    \path (A1) edge node {$ x_1 {=} 1$} node[below]{$\set{x_1}$} (B1);
    \path (B1) edge node {$x_2 {=} 2$} node[below]{$\set{x_2}$} (C1);
    \path (C1) edge node {$x_2 {=} 1$} node[below]{$\set{x_2}$} (D1);
    \path (D1) edge node {$x_1 {=} 2$} node[below]{$\set{x_1}$} (E1);

 \draw[rounded corners] (E1) -- (9, 2.5) -- (0, 2.5) --  (A1);

    \node[rotate=90,color=gray] at (-.8, 1.6) (N1) {$\text{WI}^1_2$};
    \draw[dashed,draw=gold,rounded corners=10pt] (-.5,2.8) rectangle (9.5,0.85);

 \end{tikzpicture}
 
\caption{Simulation to increment counter $C_1$, mean cost is $\frac{5\varepsilon}{3}$ for error $\varepsilon$}
\label{fig-undec_mpg_inc_new_ptu}
\end{figure}
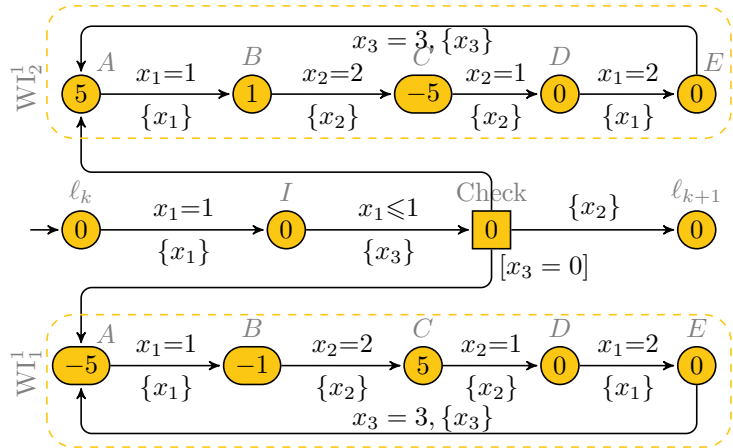
   
  \paragraph{Simulation of increment instruction}: 
   The module to
  increment $C_1$ is given in Fig.~\ref{fig-undec_mpg_inc_new_ptu}. Again,
  we start at $\ell_k$ with $x_1=\frac{1}{5^{c_1}7^{c_2}}, x_2 = 0$ and
  $x_3=0$. Let $\frac{1}{5^{c_1}7^{c_2}}$ be called as $x_{old}$.  A
  time of $1-x_{old}$ is spent at $\ell_k$.  Let the time spent at $I$ be
  denoted $x_{new}$.  To correctly increment counter 1, $x_{new}$ must
  be $\frac{x_{old}}{5}$.  No time is spent at $\text{Check}$. Player
  Max can either continue simulation of the next instruction, or can
  enter one of the widgets $\text{WI}^1_1, \text{WI}^1_2$ to verify if $x_{new}$ is indeed
  $\frac{x_{old}}{5}$.  
  
 If player Min makes an error by elapsing a time $\frac{x_{old}}{5} +\varepsilon$, then 
 player Max chooses the widget $\text{WI}^1_1$.
 It can be seen that 
 the cost of going from $A$ to $E$ once in $\text{WI}^1_1$ is $5 \varepsilon$ and the time spent in moving from $A$ to $A$
 through $E$ is 3 time units.
 Hence the mean cost 
 incurred by Min in going from $A$ to $A$ through $E$ once, 
 when Max chooses $\text{WI}^1_1$ is hence $\frac{5\varepsilon}{3}$.
 Similarly, if Min makes an error by elapsing a time $\frac{x_{old}}{5} - \varepsilon$, then 
 player Max chooses the widget $\text{WI}^1_2$. 
 The mean cost incurred in one transit from $A$ to $A$ through $E$ in $\text{WI}^1_2$  is again $\frac{5\varepsilon}{3}$.
 If Min makes no simulation error, then the mean cost incurred 
 is 0. 
 
  \paragraph{Simulation of Zero-check}:
  Figure~\ref{fig-undec_mpg_zeroCheck_ptu} shows the module for zero-check
  instruction for counter $C_2$. $\ell_k$ is a no time elapse location,
  from where player Min chooses one of the locations $\text{Check}_{c_2=0}$ or
  $\text{Check}_{c_2 \neq 0}$. Both these are player Max locations, and player Max
  can either continue the simulation, or can go to the check widgets
  $\text{W}_2^{=0}$ or $\text{W}_2^{\neq 0}$ to verify the correctness of player Min's choice.
  The widgets $\text{W}_2^{=0}$ and $\text{W}_2^{\neq 0}$ are given in
  Figure ~\ref{fig-undec_mpg_wze_ptu} and Figure \ref{fig-undec_mpg_wzne_new_pte}
  respectively.
  \begin{figure}[tbp]
\centering
    \begin{tikzpicture}[->,>=stealth',shorten >=1pt,auto,node distance=1cm,
      semithick,scale=0.8]
      \node[initial,initial text={}, player1] at (-.5,0) (A) {$0$};
      \node()[above of=A,node distance=5mm,color=gray]{$\ell_k$};
      \node () [below right of=A,node distance=6.5mm,xshift=-3mm] {$[x_3=0]$};

      \node[player2] at (2,1) (B){$0$} ;
       \node()[below of=B,node distance=5mm,color=gray]{$\text{Check}_{c_2=0}$};
       
      \node[widget] at (2,2.5) (B1){$\text{W}_2^{= 0}$ };

      \node[player2] at (2,-1) (C){$0$} ;
      \node()[above of=C,node distance=5mm,color=gray]{$\text{Check}_{c_2\neq 0}$};
      
      \node[widget] at (2,-2.5) (C1){$\text{W}_2^{\neq 0}$};

      \node[player1] at (4.5,1) (D) {$0$};
      \node()[above of=D,node distance=5mm,color=gray]{$\ell_{k+1}^1$};

      \node[player1] at (4.5,-1) (E) {$0$ };
      \node()[above of=E,node distance=5mm,color=gray]{$\ell_{k+1}^2$};

      \path (A) edge (B);
      \path (A) edge (C);
      \path (B) edge node {$x_3=0$} (B1);
      \path (B) edge node[above] {$x_3=0$} (D);
      \path (C) edge node[left] {$x_3=0$} (C1);
      \path (C) edge node[above] {$x_3=0$} (E);
    \end{tikzpicture}
    \caption{Widget $\text{WZ}_2$ simulating
      zero-check for $C_2$}

\label{fig-undec_mpg_zeroCheck_ptu}
\end{figure}
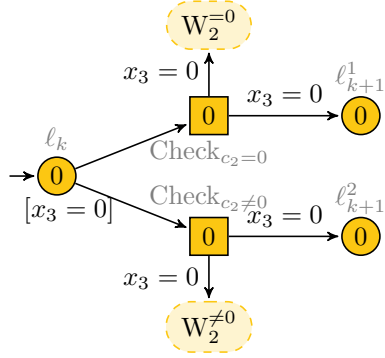

  \begin{figure}[h]
 \begin{center}
\begin{tikzpicture}[->,>=stealth',shorten >=1pt,auto,node distance=1cm,
  semithick,scale=0.8]
  \node[initial,initial text={}, player1] at (1,0) (a) {$0$};
  \node()[above of=a,node distance=5mm,color=gray]{$A$};
  
  \node[player1] at (4,0) (b) {$0$};
  \node()[above of=b,node distance=5mm,color=gray]{$B$};
  
  \node[player2] at (8,0) (chk) {$0$};
  \node()[above of=chk,node distance=5mm,color=gray]{Check};
  \node()[below right of=chk,node distance=6.5mm,xshift=1mm]{$[x_3{=}0]$};

  \node[widget] at (11,2) (wd11){$\text{WD}^1_1$};
  \node[widget] at (11,0) (wd12){$\text{WD}^1_2$};

  \node[player2] at (8,-3) (c) {$1$};
  \node()[below of=c,node distance=5mm,color=gray]{C};
  
  \node[player2] at (11,-3) (d) {$0$};
  \node()[below of=d,node distance=5mm,color=gray]{$T$};
  
  \node[player1] at (4,-3) (t) {$0$};
  \node()[below of=t,node distance=5mm,color=gray]{$F$};
  
  \path (a) edge node {$x_2{=}0$} (b);
  \path (b) edge[bend right=10] node[below] {$\set{x_3}$} (chk);
  \path (chk) edge [bend right=10] node[above] {$x_1 {\leq} 1,\{x_2\}$} (b);  
  \path (chk) edge node {} (wd11);
  \path (chk) edge node {} (wd12);
  \path (b) edge node[left] {$\begin{array}{c}x_1{=}1\\ \land x_2=0\end{array}$} (t);
  \path (t) edge [loop left] node {$x_2=0$} (t);
  \path (chk) edge node[left]{$x_1{>}1$} node[right]{$\set{x_2}$} (c);
  \path (c) edge[bend right=20] node[above] {$x_2=1$} (d);
  \path (c) edge[bend right=20] node[below] {$\set{x_2}$} (d);
  \path (d) edge[bend right=20] node[above] {$x_2=0$} (c);
\end{tikzpicture}
  \end{center}
 \caption{ Widget $\text{W}_2^{=0}$}
\label{fig-undec_mpg_wze_ptu}
\end{figure}
\begin{figure}[h]
\begin{center}
\begin{tikzpicture}[->,>=stealth',shorten >=1pt,auto,node distance=1cm,
  semithick,scale=0.8]
  \node[initial,initial text={}, player1] (a) {$0$};
  \node()[above of=a,node distance=5mm,color=gray]{$A$};
  
  \node[player1] at (4,0) (b) {$0$};
  \node()[above of=b,node distance=5mm,color=gray]{$B$};
  
  \node[player2] at (8,0) (chk) {$0$};
  \node()[above of=chk,node distance=5mm,color=gray]{$\text{Check}$};
  \node()[below of=chk,node distance=5mm]{$[x_3{=}0]$};

  \node[widget] at (11,0) (wd11){$\text{WD}^1_1$};
  \node[widget] at (11,2) (wd12){$\text{WD}^1_2$};

  \node[player2] at (0,-3) (d) {$1$};
  \node()[above of=d,node distance=5mm,color=gray]{$D$};
  
  \node[player2] at (0,-5) (t) {$0$};
  \node()[below of=t,node distance=5mm,color=gray]{$T$};
  \node[player1] at (4,-3) (t1) {$0$};
  \node()[below of=t1,node distance=5mm,color=gray]{$F$};

  \path (a) edge node {$x_2{=}0$} (b);
  \path (b) edge[bend right=10] node[below] {$x_1< 5, \set{x_3}$} (chk);
  \path (chk) edge [bend right=10] node[above] {$\{x_2\}$} (b);  
  \path (chk) edge node {} (wd11);
  \path (chk) edge node {} (wd12);
  \path (b) edge node[left] {$\begin{array}{c}x_1{=}1 \land\\  x_2=0\end{array}$} (d);
  \path (d) edge[bend left=70] node[left] {$x_2{=}1$} (t);
  \path (d) edge[bend left=70] node[right] {$\set{x_2}$} (t);
  \path (t) edge[bend left=70] node[left] {$x_2=0$} (d);
  
  \path (b) edge node[right]{$\begin{array}{c}x_1{>}1\land \\ x_2=0\end{array}$} (t1);
  \path (t1) edge [loop right] node {$x_2=0$} (t1);
\end{tikzpicture}
\caption{ Widget $\text{W}_2^{\neq 0}$}
\label{fig-undec_mpg_wzne_new_pte}
\end{center}
\end{figure}
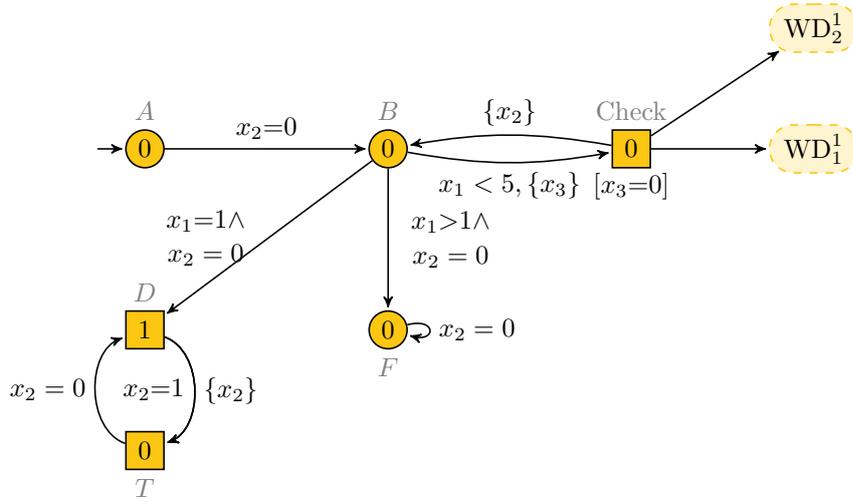
               
  \begin{itemize}
  \item Consider the case when player Min guessed that $C_2$ is zero,
    and entered the location $\text{Check}_{c_2=0}$ in
    Figure \ref{fig-undec_mpg_zeroCheck_ptu}.  Let us assume that player Max
    verifies player Min's guess by entering $\text{W}_2^{=0}$
    (Figure \ref{fig-undec_mpg_wze_ptu}).  No time is spent in the initial
    location $A$ of $\text{W}_2^{=0}$.  We are therefore at $B$ with
    $x_1=\frac{1}{5^{c_1}7^{c_2}}=x_{old}$ and $x_2,x_3=0$.  In case
    $c_1=c_2=0$, we can directly go to the $F$ state, and keep looping there
    forever, incurring mean cost 0.  If that is not the case, player Min has
    to prove his claim right, by multiplying $x_1$ with 5 repeatedly,
    till $x_1$ becomes 1; clearly, this is possible iff $c_2=0$.  The
    loop between $B$ and $\text{Check}$ precisely does this: each time
    player Min spends a time $x_{delay}$ in $B$, player Max can verify that
    $x_{delay}=4 x_{old}$ by going to $\text{WD}^1_1$ or $\text{WD}^1_2$ or come back to
    $B$.  No time is elapsed in $\text{Check}$. Finally, if $x_1=1$,
    we can go to $F$, and player Min achieves his objective. However, if
    $C_2$ was non-zero, then $x_1$ will never reach 1 after repeatedly
    multiplying $x_1$ with 5; in this case, at some point, the edge
    from $\text{Check}$ to $C$ will be enabled. In this case, the
    infinite loop between $C$ and $T$, 
   will lead to a mean cost of 1.

  \item
  Consider now the case when player Min guessed that $C_2$ is
    non-zero, and hence entered the location $\text{Check}_{c_2\neq
      0}$ in Fig.~\ref{fig-undec_mpg_zeroCheck_ptu}.  Let us assume now
    that player Max enters $\text{W}_2^{\neq 0}$
    (Fig.~\ref{fig-undec_mpg_wzne_new_pte}) to verify player Min's guess. 
    Similar to $\text{W}_2^{=0}$, no time is spent at location
    $A$ of $\text{W}_2^{\neq 0}$, and the clock values at $B$ are
    $x_1=\frac{1}{5^{c_1}7^{c_2}}=x_{old}$ and $x_2,x_3=0$.  
    If $c_1=c_2=0$, then $x_1=1$, in which case, the location $D$ is
    reached, from where, the loop between $D,T$ is taken 
    incurring a mean cost of 1.
	If that is not the case, player $\mMIN$
	repeatedly multiplies $x_1$ by 5 using the
	loop between $B$ and $\text{Check}$.
	$x_1$ becomes $1$ iff $c_2=0$.
	Once $x_1$ becomes $1$, the edge
    from $B$ to $D$ will be enabled.
    In this case, the
    infinite loop between $D$ and $T$, 
   will lead to a mean cost of 1.
   Note that if once $x_1$ becomes 1, player $\mMIN$ 
   could also wait in $B$ and transit to $\text{Check}$.
   However, due to the guard on the edge from $B$ to $\text{Check}$,
   the delay at $B$ will be less than $4x_{old}$,
   where $x_{old} = 1$.
   Say the delay is $4x_{old} - \varepsilon$
   in which case too player $\mMAX$ can go to $\text{WD}^1_1$ or $\text{WD}^1_2$ 
   and the mean cost will be greater than 0.
    If $C_2$ was non-zero, then $x_1$ will never reach 1 after repeatedly
    multiplying $x_1$ with 5; in this case, at some point, 
	$x_1$ will be greater than 1 and
    the edge
    from $B$ to $F$ will be enabled and player $\mMIN$
    can achieve its objective by moving to $F$.
      \end{itemize}
  
  \paragraph{Correctness of the construction}
  On entry into the location $\location_n$ (for HALT instruction), we
  reset clock $x_1$ to 0; from $\location_n$, we go to a state $F$
  with price 1, with a self loop that checks $x_1=1$, and resets $x_1$.
  \begin{enumerate}
  \item Assume that the two counter machine halts. If player Min
    simulates all the instructions correctly, he will incur a mean cost
    $>0$, by either reaching the $F$ after $\location_n$ :
     when player Min does not cheat, player Max has no incentive to enter any of the check widgets,
     he will just let the computation continue, till the Halt  location is reached. This will incur a mean cost 
     $>0$. 
     If player Min makes an error in his computation, player
    Max can always enter an appropriate widget, making the mean cost $>0$.   In summary, if the two counter machine halts, then  player Min has
    no strategy to achieve his goal (mean pay off $\leq 0$).
  \item Assume that the two counter machine does not halt.
    \begin{itemize}
    \item If player Min simulates all the instructions correctly, and if
      player Max never enters a check widget, then player Min incurs cost
      0, since all locations in the main modules 
      (modules simulating decrement counters,
      increment counters and zero check) have price 0. 
      Even if player Max enters some widget, the mean cost of player Min 
      is still 0, since no errors were made by Min. 
      However, if player Min makes an error, player Max can enter a check widget, ensuring that 
       the mean cost is $>0$. Thus, if the two counter machine does not halt, player Min has a strategy 
       (by making no errors) to achieve mean cost 0.
            \end{itemize}
    In summary, if the two counter machine does not halt, player Min
    has a strategy to achieve his goal, and vice-versa.
  \end{enumerate}
  Thus, player Min incurs a mean cost $\leq 0$ iff he
  chooses the strategy of faithfully simulating the two counter
  machine, when the machine does not halt. When the machine  halts, the
  mean cost incurred by player Min is more than 0
  irrespective 
  of whether he makes a simulation error or not.  \qed

  Now we argue that the $\MPG(\pta, r)$ ($r>0$) problem is undecidable for 
  PTGAs with $\geq 3$  clocks, having \emph{only 
  binary location prices and no edge prices}.
  We note that in the previous undecidability result,
  all the modules apart from the $WD^1_1, WD^1_2, WI^1_1$ and $WI^1_2$
  use binary location prices. 
  As in Figure \ref{dec_binary_neg}
and Figure \ref{dec_binary_pos}, these modules can be constructed easily. 
  The module $WD^1_1$ is chosen by the $\mMAX$ player
  when the error made by the $\mMIN$ player while
  simulating the decrement counter $C_1$ operation is $\varepsilon$.
  The module $WD^1_2$ is chosen by the $\mMAX$ player
  when the error made by the $\mMIN$ player is $-\varepsilon$.  
\end{proof}
\end{document}